\documentclass[aip,amsmath,amssymb,reprint,]{revtex4-1}

\usepackage{graphicx}% Include figure files
\usepackage{dcolumn}% Align table columns on decimal point
\usepackage{bm}
\usepackage[utf8]{inputenc}
\usepackage[T1]{fontenc}
\usepackage{mathptmx}
\usepackage{etoolbox}
\usepackage{color}

\makeatletter
\def\@email#1#2{%
 \endgroup
 \patchcmd{\titleblock@produce}
  {\frontmatter@RRAPformat}
  {\frontmatter@RRAPformat{\produce@RRAP{*#1\href{mailto:#2}{#2}}}\frontmatter@RRAPformat}
  {}{}
}%

\makeatother

\begin{document}
\preprint{AIP/123-QED}

\title{Generating magnon Bell states via parity measurement}

\author{Jia-shun Yan}
\affiliation{School of Physics, Zhejiang University, Hangzhou 310027, Zhejiang, China}

\author{Jun Jing}
\affiliation{School of Physics, Zhejiang University, Hangzhou 310027, Zhejiang, China}
\email{jingjun@zju.edu.cn}

\date{\today}

\begin{abstract}
We propose a scheme to entangle two magnon modes based on parity measurement. In particular, we consider a system that two yttrium-iron-garnet spheres are coupled to a $V$-type superconducting qutrit through the indirect interactions mediated by cavity modes. An effective parity-measurement operator that can project the two macroscopic spin systems to the desired subspace emerges when the ancillary qutrit is projected to the ground state. Consequently, conventional and multi-excitation magnon Bell states can be generated from any separable states with a nonvanishing population in the desired subspace. The target state can be distilled with a near-to-unit fidelity only by several rounds of measurements and can be stabilized in the presence of the measurement imperfection and environmental decoherence. In addition, a single-shot version of our scheme is obtained by shaping the detuning in the time domain. Our scheme that does not rely on any nonlinear Hamiltonian brings insight to the entangled-state generation in massive ferrimagnetic materials via quantum measurement.
\end{abstract}
\maketitle

\section{Introduction}

As a promising candidate for quantum control, quantum measurement is highly efficient in holding the measured system at an eigenstate or in a subspace~\cite{ZenoParadox} and steering the system to the target state~\cite{AharonovZeno,NonselectiveAharonovZeno,MeasurementEvolution,NonselectiveMeasurementEvolution,MeasurementPurification,
EntanglementPurification}. A projective measurement or postselection on the ancillary system could give rise to a positive operator-valued measure on the target system~\cite{Naimark}, which has been used to cool down a resonator~\cite{MeasurementCooling, MeasurementCoolingPRL,ExternalLevelCooling,ExpOneModeCooling,HeraldedControlMotion} to its ground state or prepare a high-ergotropy state~\cite{NonselevtiveCharging,MeasurementCharging}. A popular and well-developed projective measurement in quantum error correction~\cite{QuantumErrorCorrection,QubitInOsillator} is called parity measurement~\cite{ParityMeasurementCQED}. On mapping the parity information of the interested system to the ancillary system, parity measurement can be used to entangle double or multiple qubits~\cite{EntanglementByParityMeasurement,BellnpjQuanInf,ZenoGate}. In continuous-variable systems, schemes based on parity measurement were proposed to test the Einstein-Podolsky-Rosen state for the Bell's inequality violation~\cite{WignerEPR} and detect quantum states in the Wigner representation without use of tomographic reconstruction~\cite{DirectWigner,WignerPorbingPhase,ParityDetectionOfField}. It is not evident, however, whether or not parity measurement is efficient in generating a highly entangled state rather than an identifiable entanglement in continuous-variable systems or even macroscopic quantum systems.

Magnon attracts a significant amount of attention when such a macroscopic quantum system meets quantum information science~\cite{HybridSystem,QuantumMagnonics,CavityMagnonics}. On coupling the magnon to microwave photons~\cite{Cooperativity,HybridizingMagnonsAndPhotons,PhotonsMagnonStrongCoupling,SpinPumping,PhotonPhotonByMagnon}, mechanical phonons~\cite{CavityMagnomechanics,DynamicalBackactionMagnomechanics}, and superconducting qubits~\cite{MagnonQubitCoupling,MagnonMeetQubit,ResolvingQuanta}, the hybrid magnonic systems become controllable platforms for studying macroscopic nonclassical states and entangled states that are potential resources for diverse quantum technologies. Conventional schemes for generating entanglement of magnonic systems centred around nonlinear Hamiltonian or external nonlinear effect. A microwave field in the squeezed vacuum state can be used to prepare an entangled magnon pair in a common cavity~\cite{EntanglementFerriteSamples} or across two cavities~\cite{TwoMagnonEntanglement}. Magnon Kerr effect~\cite{KerrMagnonEntanglement}, magnetostrictive effect~\cite{MagnetostrictiveMagnonEntanglement}, magneto-optical effects~\cite{RemoteMagnonEntanglement}, and anti-ferromagnetic couplings~\cite{EnhanceMagnonMagnon,MagnonMagnonInCavity} are also meaningful to witness magnon entanglement measured by logarithmic negativity. Recently, a single-photon state is distilled from the unwanted vacuum and two-photon components with parity measurement, which is induced by detecting a desired atomic state~\cite{SinglePhotonByParityProjection}. That work inspires us to create magnon Bell states by filtering out the populations in subspaces with a distinct parity from the target state.

In this work, we transform two magnon modes from separable states to entangled states via effective parity measurement. In our system, two yttrium-iron-garnet (YIG) spheres (macroscopic spin systems) and a $V$-type qutrit are placed in a common two-mode cavity. Each cavity mode interacts individually with one magnon mode and one of the qutrit transitions, that builds up the effective coupling between magnons and the qutrit in the dispersive regime. Repeatedly projecting the ancillary qutrit onto its ground state induces parity measurement on the magnon modes. If the initial state of the magnon modes has a nonvanishing population in the subspace with a desired parity, then the induced parity measurement can create a magnon Bell state. Our scheme demonstrates robustness against measurement noise and environmental decoherence in preparing and stabilizing the Bell state with a high fidelity. Also it can be optimized to be a single-shot version adapting to a limited lifetime of the magnons.

The rest of this paper is structured as follows. In Sec.~\ref{Sec:Model} A, we briefly recall the mechanism about preparing a qubit Bell state by parity measurement. And in Sec.~\ref{Sec:Model} B, we provide a detailed derivation about the indirect couplings between the magnons and the ancillary qutrit mediated by cavity modes. Then an effective parity-measurement operator is constructed on the magnon modes. In Sec.~\ref{Sec:BellState} A and Sec.~\ref{Sec:BellState} B, a magnon Bell state is generated from a separable superposed state and a separable coherent state, respectively. In Sec.~\ref{Sec:SingleShot}, we present a single-shot measurement scheme. Finally, we summarize the whole paper in Sec.~\ref{Sec:Conclusion}.

\section{Theoretical framework}\label{Sec:Model}

Commutative parity-measurement operators find popular applications in quantum error correction to project the state of multiple qubits onto the code space~\cite{QuantumErrorCorrection}. And one of the approaches for parity detection is mapping the parity information of data qubits onto an ancillary qubit, which is readout through projective measurements~\cite{ParityMeasurementCQED}. This section is divided into two parts. We first introduce the ancillary qubit-based parity detection [see Fig.~\ref{Fig:Model}(a)] and then generalize the idea to a hybrid magnonic system [see Fig.~\ref{Fig:Model}(b)] that an effective magnon-qutrit interaction can be induced in the dispersive regime.

\subsection{Qubit-based parity detection}

\begin{figure}[htbp]
\centering
\includegraphics[width=0.9\linewidth]{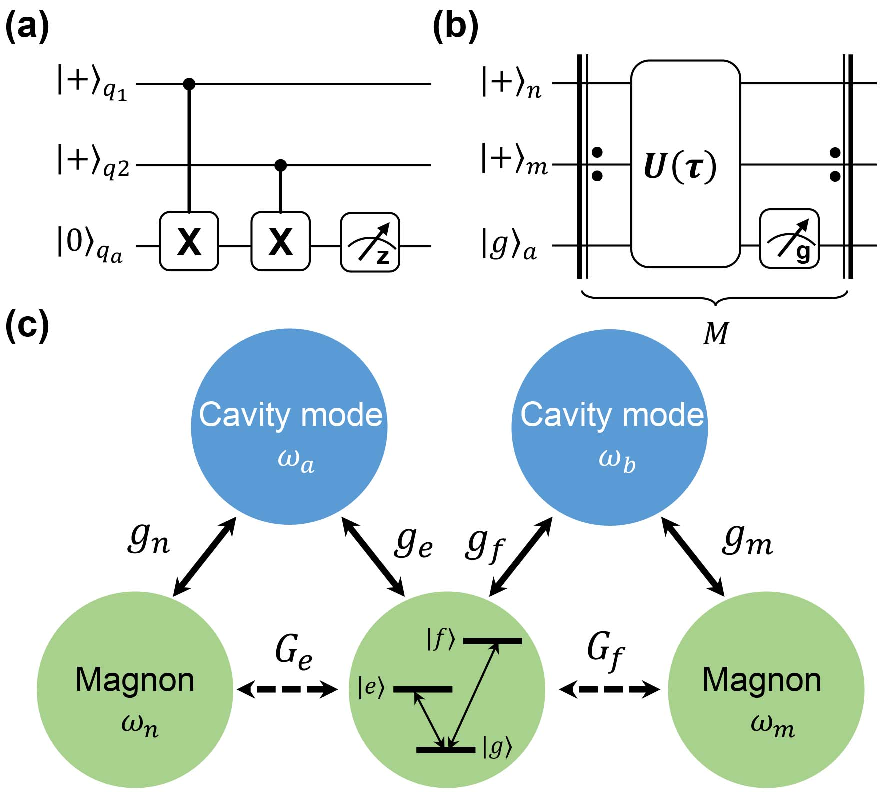}
\caption{(a) Diagram for a typical ancillary-qubit-based parity measurement that gives rise to a qubit Bell state. (b) Diagram for our scheme to obtain a magnon Bell state by repeated rounds of free evolution and projective measurement. (c) Two magnon modes $n$ and $m$ are individually coupled to two cavity modes $a$ and $b$ with coupling strengths $g_n$ and $g_m$, respectively. The transitions $|g\rangle\leftrightarrow|e\rangle$ and $|g\rangle\leftrightarrow|f\rangle$ in a $V$-type three-level system are coupled to the cavity mode-$a$ and mode-$b$ with coupling strengths $g_e$ and $g_f$, respectively. In the dispersive regime, magnon mode-$n$ and mode-$m$ are therefore indirectly coupled to $|g\rangle\leftrightarrow|e\rangle$ and $|g\rangle\leftrightarrow|f\rangle$ with effective coupling strengths $G_e$ and $G_f$, respectively.}\label{Fig:Model}
\end{figure}

As shown in Fig.~\ref{Fig:Model}(a), a Bell state $|\Phi^+\rangle=(|00\rangle+|11\rangle)/\sqrt{2}$ of two transmon qubits (data qubits) could be generated by mapping the parity information to the ancillary qubit via two controlled NOT (CNOT) gates and one projective measurement on the initial state of the ancillary qubit~\cite{BellnpjQuanInf}. The data qubits $q_1$ and $q_2$ are prepared in a superposed state $|+\rangle|+\rangle$ with $|+\rangle\equiv(|0\rangle+|1\rangle)/\sqrt{2}$ and the ancillary qubit $q_a$ is prepared in $|0\rangle$. Applying two sequential CNOT gates $C(q_i,q_a)$, where $q_i$, $i=1,2$, is the control qubit and $q_a$ is the target one, on the system, one can obtain an entangled state that correlates the parity information of data qubits to the ancillary qubit:
\begin{equation}
C(q_1,q_a)C(q_2,q_a)|++\rangle|0\rangle_{q_a}=|\Phi^+\rangle|0\rangle_{q_a}+|\Psi^+\rangle|1\rangle_{q_a}.
\end{equation}
Here $|\Psi^+\rangle=(|01\rangle+|10\rangle)/\sqrt{2}$ is another Bell state with the same XX parity as $|\Phi^+\rangle$, i.e., $\langle\sigma_x\otimes\sigma_x\rangle=1$ for both states. Therefore, measuring the ancillary qubit and confirming it is in the initial state $|0\rangle_{q_a}$ heralds that the data qubits are in the target Bell state $|\Phi^+\rangle$.

In the circuit, the two CNOT gates and the projective measurement on the ancillary qubit constitute a nonunitary operator for the data qubits. It reads,
\begin{equation}\label{ParityProjection}
\mathcal{M}=\langle 0|C(q_1,q_a)C(q_2,q_a)|0\rangle_{q_a}=|00\rangle\langle 00|+|11\rangle\langle 11|,
\end{equation}
which is a parity-measurement operator projecting the two-qubit system into the subspace with a special parity $\langle \sigma_z\otimes\sigma_z\rangle=1$. Then the Bell-state generation could be equivalently described by applying a parity measurement on the initial state, i.e., $|\Phi^+\rangle=\mathcal{M}|++\rangle/P$, where $P=\langle++|\mathcal{M}|++\rangle$ is the measurement probability.

\subsection{Cavity-mediated magnon-qutrit coupling and effective parity measurement}\label{Subsec:EffectiveOperator}

An effective nonunitary operator similar to Eq.~(\ref{ParityProjection}) for magnon modes can be constructed by the unitary evolution of the whole system and the projective measurement on the ancillary qutrit. As shown in Fig.~\ref{Fig:Model}(c), our model consists of two cavity modes, two magnon modes, and a $V$-type qutrit. The cavity mode-$a$ is coupled to the magnon-$n$ and the qutrit transition $|g\rangle\leftrightarrow|e\rangle$ and cavity mode-$b$ is coupled to the magnon-$m$ and the qutrit transition $|g\rangle\leftrightarrow|f\rangle$. The crosstalk between magnon-$n$ and cavity mode-$b$ and that between magnon-$m$ and cavity mode-$a$ are assumed to be negligible~\cite{MagnonXu}. The full Hamiltonian can be written as ($\hbar\equiv1$)
\begin{equation}
\begin{aligned}
H=&\sum_{l=a,b,n,m}\omega_ll^\dagger l+\sum_{i=e,f}\omega_i |i\rangle\langle i|\\
&+g_n(a^\dagger n + an^\dagger) + g_e(a^\dagger\sigma_{eg}^-+a\sigma_{eg}^+)\\
&+g_m(b^\dagger m + bm^\dagger)+ g_f(b^\dagger\sigma_{fg}^-+b\sigma_{fg}^+)
\end{aligned}
\end{equation}
where $\omega_{a,b}$ are the cavity-mode frequencies, $\omega_{n,m}$ are the magnon-mode frequencies, $\omega_{e,f}$ represent frequencies of excited levels in qutrit, and the ground-state frequency of the qutrit is set to be zero. $n$ and $m$ ($a$ and $b$) are annihilation operators for magnon modes (cavity modes). $\sigma_{ig}^-\equiv|g\rangle\langle i|$ and $\sigma_{ig}^+\equiv|i\rangle\langle g|$, $i=e,f$, are qutrit transition operators between excited states and the ground state. $g_{n,m}$ and $g_{e,f}$ represent the interaction strengths of photon-magnon coupling and photon-qutrit coupling, respectively. In practice, the interactions between qutrit and cavity modes can be realized by placing the superconducting qutrit near a common electric-field antinode of the two modes; and two pairs of magnon-photon interactions can be realized by mounting the two YIG spheres near two distinct magnetic-field antinodes of cavity modes. An alternative configuration relies on a cross-shaped cavity with two crossing dispersive microwave modes~\cite{TwoMagnonEntanglement,EntanglementByDressedField}, where the qutrit is placed at the intersection of two modes and the two YIG spheres are individually placed in between the intersection and the end mirrors of the two cavity branches. The detunings between magnons (level splitting of the qutrit) and the coupled cavity modes are labelled with $\Delta_n=\omega_n-\omega_a$ and $\Delta_m=\omega_m-\omega_b$ ($\Delta_e=\omega_e-\omega_a$ and $\Delta_f=\omega_f-\omega_b$), respectively. In the dispersive regime that these detunings are sufficiently larger than the interaction strengths, i.e., $|g_{i}/\Delta_i|\ll 1$, the qutrit-magnon couplings could be induced by the photon-magnon and photon-qutrit couplings~\cite{CircuteQED}. In this case, both magnon modes and qutrit transitions are far off-resonant from the relevant cavity modes that are only virtually populated. According to the Schrieffer-Wolff transformation~\cite{SWtransformation}, in a rotating frame with respect to
\begin{equation}
\begin{aligned}
S&=\frac{g_n}{\Delta_n}(an^\dagger-a^\dagger n)+\frac{g_m}{\Delta_m}(bm^\dagger - b^\dagger m)\\
&+\frac{g_e}{\Delta_e}(a\sigma_{eg}^+ - a^\dagger \sigma_{eg}^-)+\frac{g_f}{\Delta_f}(b\sigma_{fg}^+-b^\dagger\sigma_{fg}^-),
\end{aligned}
\end{equation}
an effective Hamiltonian to the second order of $g_i/\Delta_i$ can be obtained by the Baker-Campbell-Hausdorff expansion, which reads
\begin{equation}\label{HeffComplete}
\begin{aligned}
\tilde{H}&=e^{-S}He^S=\sum_{l=a,b,n,m}\tilde{\omega}_ll^\dagger l+\sum_{i=e,f}\tilde{\omega}_i |i\rangle\langle i| \\
&+ G_e(n\sigma_{eg}^++n^\dagger\sigma_{eg}^-)+ G_f(m\sigma_{fg}^++m^\dagger\sigma_{fg}^-) \\
&+ \chi_ea^\dagger a\sigma^z_{eg}+\chi_fb^\dagger b\sigma^z_{fg}+ G_{fe} (a^\dagger b\sigma_{fe}^++ab^\dagger\sigma_{fe}^-)
\end{aligned}
\end{equation}
with $\sigma^z_{ij}\equiv|i\rangle\langle i|-|j\rangle\langle j|$. Here the tilde frequencies
\begin{equation}
\begin{aligned}
&\tilde{\omega}_a=\omega_a-\chi_n, \quad \tilde{\omega}_b=\omega_b-\chi_m, \quad \tilde{\omega}_n=\omega_n+\chi_n,\\
&\tilde{\omega}_m=\omega_m+\chi_m, \quad \tilde{\omega}_e=\omega_e+\chi_e, \quad \tilde{\omega}_f=\omega_f+\chi_f,
\end{aligned}
\end{equation}
include the Lamb shifts $\chi_i=g_i^2/\Delta_i$ and the cavity-induced magnon-qutrit couplings take the form of
\begin{equation}
G_e=\frac{g_eg_n}{2}\left(\frac{1}{\Delta_e}+\frac{1}{\Delta_n}\right),\quad G_f=\frac{g_mg_f}{2}\left(\frac{1}{\Delta_f}+\frac{1}{\Delta_m}\right).
\end{equation}
Note the last term in Eq.~(\ref{HeffComplete}) is a three-body interaction about the two cavity modes and the qutrit transitions $|e\rangle\leftrightarrow|f\rangle$. If the cavity modes do not significantly deviate from the initial vacuum states, i.e., $\langle a^\dagger a\rangle,\langle b^\dagger b\rangle\approx0$, then all the three terms of the last line in Eq.~(\ref{HeffComplete}) can be ignored and the effective Hamiltonian can be rewritten as
\begin{equation}
\begin{aligned}
&\tilde{H}\approx\tilde{\omega}_aa^\dagger a+\tilde{\omega}_bb^\dagger b+\tilde{\omega}_nn^\dagger n+\tilde{\omega}_m m^\dagger m+\tilde{\omega}_e|e\rangle\langle e|\\
&+\tilde{\omega}_f|f\rangle\langle f|+ G_e(n\sigma_{eg}^++n^\dagger\sigma_{eg}^-) + G_f(m\sigma_{fg}^++m^\dagger\sigma_{fg}^-).
\end{aligned}
\end{equation}
In the rotating frame with respect to $H_R=\tilde{\omega}_aa^\dagger a + \tilde{\omega}_b b^\dagger b+ \tilde{\omega}_n(n^\dagger n+|e\rangle\langle e|)+\tilde{\omega}_m(m^\dagger m+|f\rangle\langle f|)$, a Jaynes-Cummings-like (JC) Hamiltonian emerges:
\begin{equation}\label{Ham}
\begin{aligned}
H_{\rm eff}=&\tilde{\Delta}_e|e\rangle\langle e|+\tilde{\Delta}_f|f\rangle\langle f|+G_e(n\sigma_{eg}^++n^\dagger\sigma_{eg}^-)\\
 +& G_f(m\sigma_{fg}^++m^\dagger\sigma_{fg}^-),
\end{aligned}
\end{equation}
where $\tilde{\Delta}_e=\tilde{\omega}_e-\tilde{\omega}_n$ and $\tilde{\Delta}_f=\tilde{\omega}_f-\tilde{\omega}_m$. Here the effective couplings between magnons and qutrit transitions are induced by exchanging the virtual photons of cavity modes. In the recent experiments, for a $1$ mm-diameter YIG sphere with a bare frequency $\omega_{n,m}\sim$ GHz, the dispersive coupling strengths $G_{e,f}$ are in order of $10$ MHz~\cite{MagnonXu,xuBellstate}. In Appendix~\ref{AppendSec:OneModeCoupling}, we provide an alternative model with a single cavity mode to achieve the same effective Hamiltonian as in Eq.~(\ref{Ham}).

An effective parity-measurement operator can be induced by an evolution-and-measurement cycle. Initially, the qutrit is in its ground state $|g\rangle$ and the state of two magnon modes is separable $\rho(0)=\rho_n(0)\otimes\rho_m(0)$ with a non-vanishing overlap with the target state, then the initial state of the whole system is $\rho_{\rm tot}(0)=|g\rangle\langle g|\otimes\rho(0)$. After a period of joint evolution by $U(\tau)=\exp(-iH_{\rm eff}\tau)$, the qutrit is measured by a projective operator $M_g=|g\rangle\langle g|$ and then the whole system becomes
\begin{equation}
\rho_{\rm tot}(\tau)=\frac{M_gU(\tau)\rho_{\rm tot}(0)U^\dagger(\tau)M_g}{{\rm Tr}[M_gU(\tau)\rho_{\rm tot}(0)U^\dagger(\tau)M_g]}.
\end{equation}
According to Naimark's dilation theorem~\cite{Naimark}, the projection applied on the ancillary qutrit induces a positive operator-valued measure $\mathcal{M}(\tau)[\mathcal{O}]=V_g(\tau)\mathcal{O}V_g^\dagger(\tau)$ on the magnon modes. Then the magnon state can be expressed as
\begin{equation}
\rho(\tau)=\frac{V_g(\tau)\rho(0)V_g^\dagger(\tau)}{P_g},
\end{equation}
where $V(\tau)\equiv\langle g|U(\tau)|g\rangle$ is a nonunitary evolution operator acting on the magnon space and $P_g={\rm Tr}[V_g(\tau)\rho(0)V_g^\dagger(\tau)]$ is the measurement probability. Assuming that the detunings are the same $\tilde{\Delta}_e=\tilde{\Delta}_f=\Delta$, the measurement-induced evolution operator takes the form of
\begin{equation}\label{KrausOperator}
V_g(\tau)=e^{-i\Delta\tau/2}\sum_{n,m\geq0}\alpha_{nm}(\tau)|nm\rangle\langle nm|
\end{equation}
with coefficients
\begin{equation}\label{coefficient}
\alpha_{nm}(\tau)=\cos\left(\Omega_{nm}\tau\right)+i\frac{\Delta}{2\Omega_{nm}}\sin\left(\Omega_{nm}\tau\right),
\end{equation}
where $\Omega_{nm}=(G_e^2n+G_f^2m+\Delta^2/4)^{1/2}$ is the Rabi frequency. Note that there is no off-diagonal element for the operator in Eq.~(\ref{KrausOperator}) and the coefficients satisfy $|\alpha_{nm}(\tau)|\leq1$. It means that $V_g(\tau)$ acts as a population-filtering operator. Populations on the special states with $|\alpha_{nm}(\tau)|=1$ will be conserved and those on the other states with $|\alpha_{nm}(\tau)|<1$ will be gradually eliminated by repeating such evolution-and-measurement rounds.

It is straightforward to see that the ground state $|00\rangle$ always satisfies $|\alpha_{00}|=1$ due to the fact that it is decoupled from the time evolution. Then to generate a Bell state $|\Phi^+\rangle=(|00\rangle+|11\rangle)/\sqrt{2}$, the effective coupling strengths $\Omega_{e,f}$, the detuning $\Delta$, and the free-evolution interval $\tau$ should be so engineered that $|\alpha_{11}(\tau)|=1$ with $\tau=2k\pi/\Omega_{11}$, where $k\in\mathbb{N}_+$. With $k=1$ or $\tau=\tau_0\equiv2\pi/\Omega_{11}$, we have
\begin{equation}\label{Vgtau0}
\begin{aligned}
V_g(\tau_0)&=|00\rangle\langle 00|+e^{-i\phi}|11\rangle\langle 11|+e^{-i\phi}\biggl[\alpha_{01}(\tau_0)|01\rangle\langle 01|\\&+\alpha_{10}(\tau_0)|10\rangle\langle 10|+\sum_{n+m\geq2}\alpha_{nm}(\tau_0)|nm\rangle\langle nm|\biggr]
\end{aligned}
\end{equation}
with $\phi=\pi\Delta/\Omega_{11}$. Then after $M$ rounds of free-evolution and the ground-state projection on the qutrit, the magnon state becomes
\begin{equation}
\begin{aligned}
\rho(M\tau_0)&=\underbrace{\mathcal{M}(\tau_0)\circ\mathcal{M}(\tau_0)\circ\cdots\circ\mathcal{M}(\tau_0)}_{M}[\rho(0)]\\
&=V_g^M(\tau_0)\rho(0)V_g^{\dagger M}(\tau_0)/P_s
\end{aligned}
\end{equation}
with
\begin{equation}\label{ParityOperator}
\begin{aligned}
V_g^M(\tau_0)&=|00\rangle\langle00|+e^{-iM\phi}|11\rangle\langle11|\\
&+e^{-iM\phi}\biggl[\alpha_{01}^M(\tau_0)|01\rangle\langle01|+\alpha_{10}^M(\tau_0)|10\rangle\langle10|\\
&+\sum_{n+m\geq2}\alpha_{nm}^M(\tau_0)|nm\rangle\langle nm|\biggr]
\end{aligned}
\end{equation}
and a success probability $P_s\equiv{\rm Tr}[V_g^M(\tau_0)\rho(0)V_g^{\dagger M}(\tau_0)]$. As measurements are repeated, the absolute value of the coefficients $|\alpha_{01}|^M$ and $|\alpha_{10}|^M$ in $V_g^M$ will exponentially vanish due to $|\alpha_{10}|,|\alpha_{01}|<1$. And the same thing occurs for $|\alpha_{02}|^M$ and $|\alpha_{20}|^M$ when $G_e\neq G_f$. Assuming that the two magnon modes are near-resonant to the relevant qutrit transitions and none of them is double excited, the nonunitary evolution operator becomes an effective projection operator
\begin{equation}\label{EffectiveParityOperator}
\tilde{V}_g^M(\tau_0)\approx|00\rangle\langle 00|+|11\rangle\langle 11|
\end{equation}
in the same form as Eq.~(\ref{ParityProjection}). Note in our case, $|0\rangle$ and $|1\rangle$ represent the number states of magnon mode. For a resonator system, the parity operator is defined as $Q_l\equiv e^{i\pi l^\dagger l}$, which is widely used in Wigner tomography~\cite{DirectWigner,WignerPorbingPhase}. Then a completed parity operator of two magnon modes is written as
\begin{equation}
Q_{nm}=Q_n\otimes Q_m=\sum_{n,m=0}(-1)^{n+m}|nm\rangle\langle nm|,
\end{equation}
where $|00\rangle$ and $|11\rangle$ are in the even parity subspace such that $\langle Q_{nm}\rangle=1$ and $|01\rangle$ and $|10\rangle$ are in the odd parity subspace such that $\langle Q_{nm}\rangle=-1$. It means that our parity-measurement operator obtained in Eq.~(\ref{EffectiveParityOperator}) is actually a {\it partial} parity operator, projecting the magnon system into the even-parity magnon subspace.

In the presence of detection noise, the measurement operator could be modified to $M_{\tilde{g}}=|\tilde{g}\rangle\langle\tilde{g}|$ with $|\tilde{g}\rangle=|g\rangle+\epsilon_e|e\rangle+\epsilon_f|f\rangle$, where the deviation ratios are assumed as $\epsilon_e=\epsilon_f=\epsilon$ for simplicity. To the first order of $\epsilon$, the effective measurement operator in Eq.~(\ref{EffectiveParityOperator}) is found to be
\begin{equation}\label{VNoise}
\begin{aligned}
\tilde{V}_{\tilde{g}}^M&\approx|00\rangle\langle00|+|11\rangle\langle 11|+\epsilon M\bigg[\mu_{10}|00\rangle\langle10|+\nu_{01}|00\rangle\langle01|\\
&+\mu_{11}|11\rangle\langle01|+\nu_{11}|11\rangle\langle10|\bigg].
\end{aligned}
\end{equation}
Calculation details are provided in Appendix~\ref{AppendSec:OperatorWithNoise}. The noise terms in Eq.~(\ref{VNoise}) could transfer the populations on $|01\rangle$ and $|10\rangle$ to $|00\rangle$ and $|11\rangle$ yet with distinct rates, i.e., $\mu_{10},\nu_{01}\neq\mu_{11},\nu_{11}$. They will break the population balance between $|00\rangle$ and $|11\rangle$ and weaken the parity-measurement effect in the end.

\section{Preparing Bell states}\label{Sec:BellState}

Generating entanglement between macroscopic systems is an ongoing effort in quantum science, which facilitates quantum-enhanced sensing and exploration of the fundamental limits of quantum theory~\cite{MechanicalAndSpinEntanglement}. Nevertheless, it is subject to the precise control over the system and the stability under the environmental influence~\cite{MacroscopicEntanglement}. In this section, the parity measurement with the operator in Eq.~(\ref{EffectiveParityOperator}) is used to generate a magnon Bell state of two YIG spheres. We first consider an ``easy-mode'' case to generate a conventional Bell state where the magnons are initialized as separable single-excitation superposed states. And then in an open-quantum-system scenario, we check the robustness of our measurement-based scheme against the environmental decoherence. To generate double- and even multiple-excitation Bell states of two magnon modes, we consider that they start from separable coherent states. It is a ``hard-mode'' case for quantum control with more undesired populations over other subspaces.

\subsection{Preparation Bell states from separable single-magnon superposed state}\label{superposed}

\begin{figure}[htbp]
\centering
\includegraphics[width=0.9\linewidth]{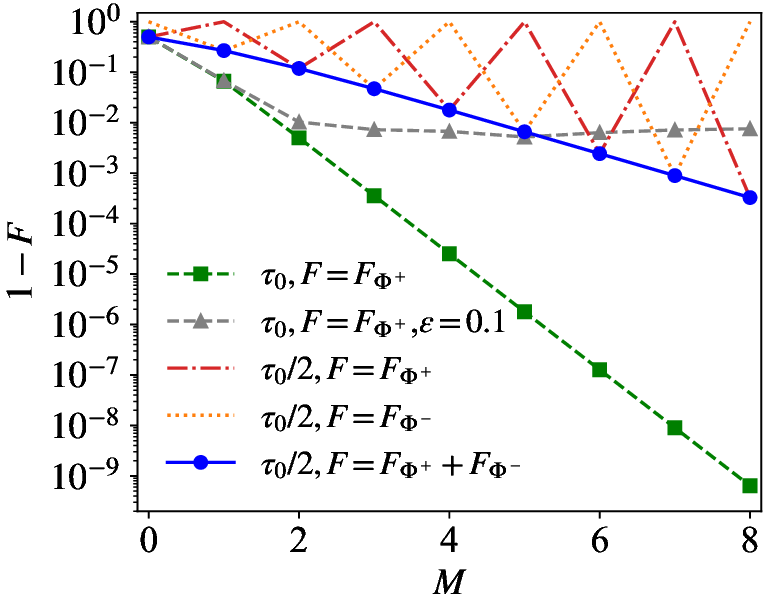}
\caption{Infidelity of the magnon system with respect to the Bell states $|\Phi^{\pm}\rangle\equiv(|00\rangle\pm|11\rangle)/\sqrt{2}$ with a measurement interval $\tau_0\equiv2\pi/\Omega_{11}$ or $\tau_0/2$. $\epsilon=0.1$ represents the upperbound of the detection noise. Magnons are assumed to be resonant with the relevant qutrit transitions $\tilde{\Delta}_e=\tilde{\Delta}_f=0$ and the effective coupling strengths are the same $G_e=G_f=10^{-3}\omega_m$.}\label{Fig:BellFidelity}
\end{figure}

We first consider preparing a magnon Bell state $|\Phi^+\rangle$ using the effective purity measurements in a system that the two magnon modes are initially at the same superposed state:
\begin{equation}
|\psi_i\rangle=\frac{|0\rangle+|1\rangle}{\sqrt{2}}\otimes\frac{|0\rangle+|1\rangle}{\sqrt{2}}.
\end{equation}
The single-excitation superposed state in magnon system has been created in a recent experiment~\cite{MagnonXu}. The Bell state $|\Phi^+\rangle$ could then be straightforwardly generated under several rounds of parity measurements, i.e., $|\Phi^+\rangle=\tilde{V}_g^M(\tau)|\psi_i\rangle/P_s$. In Fig.~\ref{Fig:BellFidelity}, the infidelity $1-F$ about the target Bell state with $F=F_{\Phi^+}\equiv\langle\Phi^+|\rho(t)|\Phi^+\rangle$ is plotted with a green dashed line marked with squares. The cavity-induced coupling strengths $G_e$ and $G_f$ are set the same for the best performance in fidelity. The reasons are illustrated in Appendix~\ref{AppendSec:OptimalCoupling}. As the rounds of evolution and measurement are repeated, the state infidelity decreases in an exponential way. When $M=8$, it is reduced by over nine orders in magnitude from the initial state. The result means that a near-to-perfect parity measurement can be induced by several projective measurements on the ancillary qutrit, conserving the initial population over the even parity subspace and filtering out undesired population over the odd parity subspace. If the measurement imperfection described by Eq.~(\ref{VNoise}) occurs in each measurement, where the deviation ratio $\epsilon$ is randomly distributed in $[0, 0.1]$, then the parity measurement becomes less effective (see the gray dashed line marked with triangles). Yet the fidelity can be maintained at a satisfactory level with $F=0.998$ when $M=5$.

When the measurement interval is set as one half of the interval $\tau=\tau_0/2$ for $|\Phi^+\rangle$, we have $\alpha_{11}(\tau)=-1$ with $n=m=1$ due to Eq.~(\ref{coefficient}). It means that the effective measurement operator in Eq.~(\ref{EffectiveParityOperator}) becomes dependent on the parity of the measurement number $M$, i.e.,
\begin{equation}
\tilde{V}_g^M(\tau)\approx|00\rangle\langle00|+(-1)^M|11\rangle\langle11|.
\end{equation}
We can therefore generate another Bell state $|\Phi^-\rangle=(|00\rangle-|11\rangle)/\sqrt{2}$ with an odd $M$, which is described by the infidelity $1-F_{\Phi^-}$ in Fig.~\ref{Fig:BellFidelity} (see the orange-dotted line). And when $\tau=\tau_0/2$ and $M$ is even, $1-F_{\Phi^+}$ approaches vanishing. So that as the measurements are repeated, the population of the two magnon modes is swapped between $|\Phi^+\rangle$ and $|\Phi^-\rangle$ (see the staggered orange-dotted and red-dot-dashed lines) and gradually concentrates in the subspace of $\{|00\rangle,|11\rangle\}$ (see the blue line marked with circles). In other words, our scheme is capable of preparing and transferring two distinct Bell states merely by controlling the parity of the number of measurements.

The parity-measurement operator is induced by projecting the ancillary qutrit. Thus our scheme is essentially nondeterministic and one of the key metrics about the scheme efficiency is the success probability $P_s$. In our model, the initial population over the target entangled state provides a lower bound for the success probability, i.e., $P_s\geq|\langle\Phi^+|\psi_i\rangle|^2$. In generating $|\Phi^{\pm}\rangle$, the success probability after $M=8$ measurements is found to be about $P_s\approx50\%$. It is consistent with the initial condition and much larger than the probability in a previous scheme~\cite{RemoteMagnonEntanglement} for generating a magnon Bell state, which was based on the magnon-induced Brillouin light scattering in an optomagnonic weak-coupling regime.

Our scheme is dramatically distinct in mechanism from those depending on introducing or inducing nonlinear Hamiltonian or interaction~\cite{EntanglementFerriteSamples,TwoMagnonEntanglement,KerrMagnonEntanglement,EntanglementByDressedField,
MagnetostrictiveMagnonEntanglement,RemoteMagnonEntanglement,EnhanceMagnonMagnon,MagnonMagnonInCavity}. For example, in Ref.~\cite{EntanglementByDressedField}, the entanglement between two magnon modes depends on the two-magnon-mode squeezing, that is generated by the anti-JC interaction through dressing the atomic transitions by two classical fields. In contrast, our scheme depends on projecting a separable state of the system into a desired parity subspace. With respect to the output, squeezing induced entanglement is quantitatively evaluated by the variances of quadratures and our scheme directly yields a particular entangled state, i.e., the Bell state.

\begin{figure}[htbp]
\centering
\includegraphics[width=0.9\linewidth]{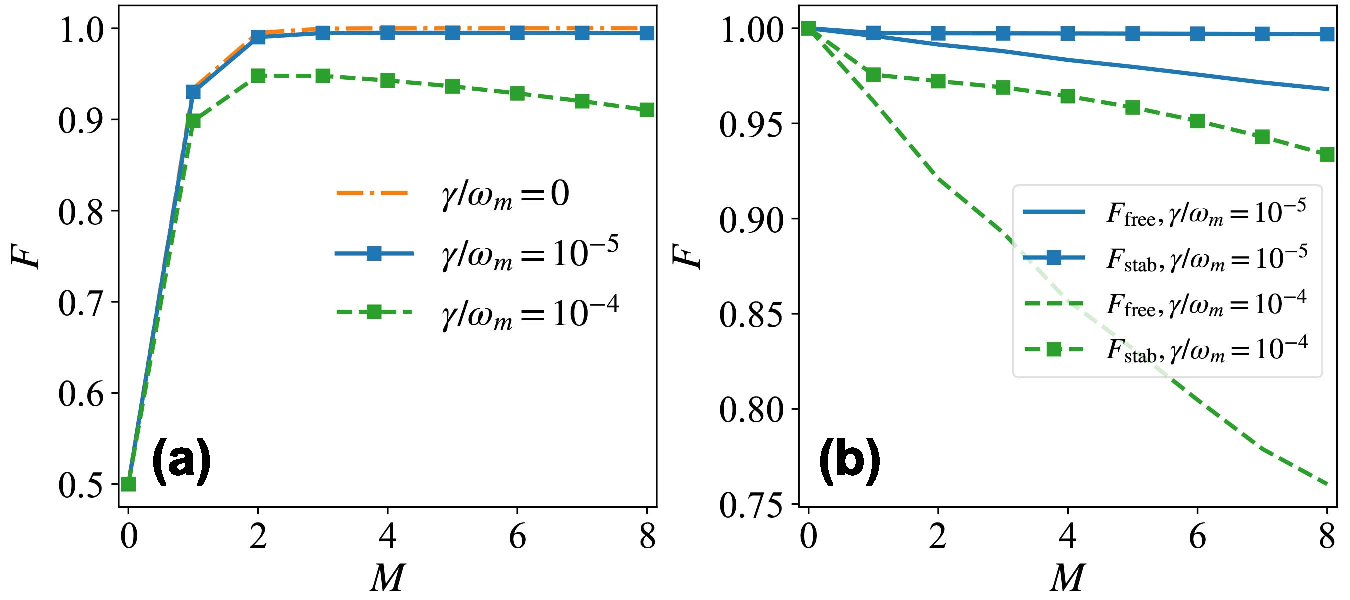}
\caption{Fidelities of the Bell state $|\Phi^+\rangle$ in (a) preparation process and (b) stabilization process under various decoherence rates. The coupling strengths are set as $G_e=G_f=6\times10^{-3}\omega_m$ and the measurement interval is set as $\tau=\tau_0$.}\label{Fig:Stablization}
\end{figure}

Our parity-measurement-based scheme can be performed in the presence of the environmental decoherence. In the Bell-state preparation process, the free evolution of the whole system between neighboring measurements is evaluated by the master equation
\begin{equation}\label{MasterEq}
\dot\rho(t)=-i[H_{\rm eff},\rho(t)]+\gamma_n\mathcal{D}[n]\rho(t)+\gamma_m\mathcal{D}[m]\rho(t),
\end{equation}
where $H_{\rm eff}$ is the effective Hamiltonian in Eq.~(\ref{Ham}) and $\mathcal{D}[A]$ represents the Lindblad superoperator
\begin{equation}
\mathcal{D}[A]\rho(t)\equiv A\rho(t)A^\dagger - \frac{1}{2}\{A^\dagger A,\rho(t)\}.
\end{equation}
For the gigahertz Kittel modes, the effective coupling between the mode and a qubit can be as great as $10$~MHz and the decay rate of the Kittel mode is about $1$~MHz~\cite{MagnonQubitCoupling,MagnonXu}. About $M=10$ measurements can therefore be performed within the magnon lifetime. In Fig.~\ref{Fig:Stablization}(a), we plot the fidelity $F=F_{\Phi^+}$ in the preparation process under various decoherence rates, where the decoherence rate for each magnon mode is assumed to be the same $\gamma_n=\gamma_m=\gamma$. It is found that the target-state fidelity still rapidly increases with measurements even under decoherence. With a comparatively large decay rate $\gamma/\omega_m=10^{-4}$, the state fidelity after $M=8$ measurements is over $F=0.91$, indicating that our preparation scheme is not fragile to the environment-induced decoherence.

The projection-induced parity measurement is also capable of stabilizing the system against decoherence when the Bell-state generation is completed. In Fig.~\ref{Fig:Stablization}(b), we compare the state fidelities under the free evolution with decoherence $F_{\rm free}$ and that under both decoherence and repeated projective measurements $F_{\rm stab}$. It is found that in the absence of measurements, the fidelity $F_{\rm free}$ monotonically decreases with time due to the magnon loss. When $t=8\tau_0$, the fidelity will be lower than $0.77$ for $\gamma/\omega_m=10^{-4}$. In contrast, the decaying tendancy of the state fidelity can be significantly suppressed by the parity measurements. For $\gamma/\omega_m=10^{-4}$, the fidelity can be held around $F=0.93$ after $M=8$ measurements; and for $\gamma/\omega_m=10^{-5}$, the fidelity is held close to one. Here the dominant error in generating the Bell state from the environmental decoherence is the single-magnon loss. Then the population on $|11\rangle$ tends to leak to the odd-parity subspace of $\{|10\rangle, |01\rangle\}$. Our parity measurement induced by the projection $\tilde{V}_g^M$ in Eq.~(\ref{EffectiveParityOperator}) could avoid this error by suppressing the leakage and projecting the system into the target subspace $\{|00\rangle, |11\rangle\}$ with even parity, which thus works as a stabilizer for entanglement preparation.

\subsection{Preparation Bell states from separable single-magnon coherent state}

\begin{figure}[htbp]
\centering
\includegraphics[width=0.9\linewidth]{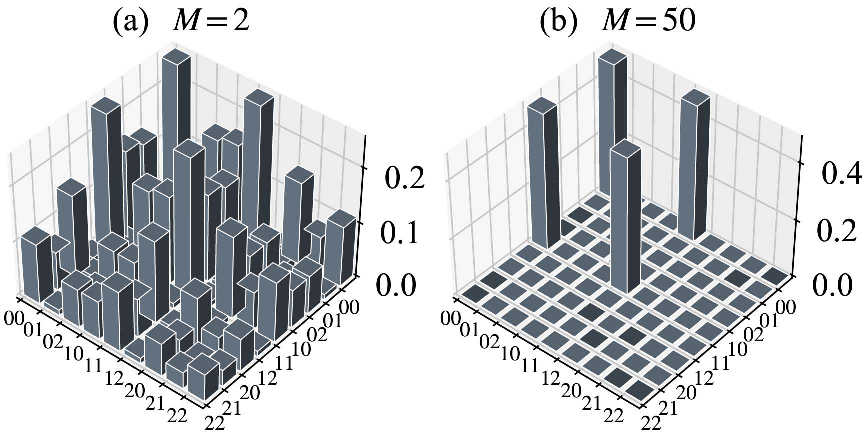}
\caption{Tomography of the two magnon modes after (a) $M=2$ and (b) $M=50$ measurements. Each mode starts from the same coherent state $\beta_n=\beta_m=1$. The effective coupling strengths are set as $G_e=1.0\times10^{-3}\omega_m$ and $G_f=1.2G_e$ and magnons are assumed to be resonant with relevant qutrit transitions, i.e., $\tilde{\Delta}_e=\tilde{\Delta}_f=0$.}\label{Fig:Coherent}
\end{figure}

We can choose a ``hard-mode'' about the initial states for Bell-state generation by parity measurement, which has a wide distribution over the Hilbert space. When the magnon modes are in their individual coherent states, the size of the subspace with distinct parity is clearly much larger than that of the prior initial state in Sec.~\ref{superposed}. Consequently, more measurements are necessary to filter out the undesired populations than that for a single-excitation superposed state in Fig.~\ref{Fig:BellFidelity}. The initial state of magnon modes is written as
\begin{equation}
|\beta\rangle=|\beta_n\rangle\otimes|\beta_m\rangle,\quad |\beta_{l=n,m}\rangle= e^{-\frac{|\beta_l|^2}{2}}\sum_j\frac{\beta^j_l}{\sqrt{j}}|j\rangle.
\end{equation}
The magnon coherent state could be readily realized by applying a microwave drive in resonance with the Kittel mode, which serves as a displacement operator on the vacuum state of the magnon mode $D(\beta_i)|0\rangle$~\cite{MagnonXu}. To generate the Bell state $|\Phi^+\rangle$, one can choose $\beta_n=\beta_m=1$ to have a significant initial overlap with the target state. We plot the state tomographies for the two magnon modes after $M=2$ and $M=50$ rounds of measurements in Fig.~\ref{Fig:Coherent}. It is found that the magnon state distribution has been dramatically reshaped only by $M=2$ measurements, where the target-state population already prevails over the others. After $M=50$ measurements, a magnon Bell state is generated with a fidelity $F\approx0.97$.

\begin{figure}[htbp]
\centering
\includegraphics[width=0.9\linewidth]{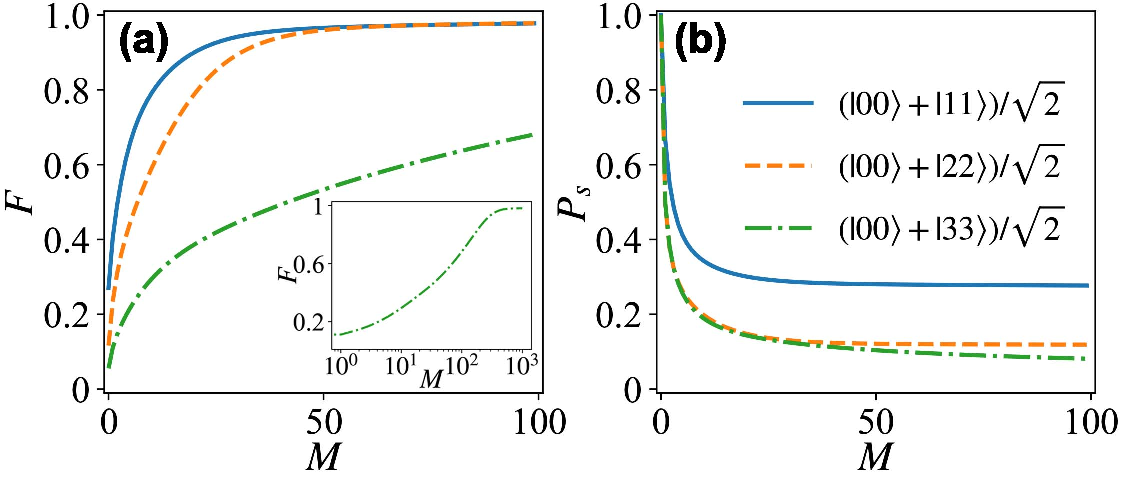}
\caption{(a) Fidelities of the general Bell state $\langle\Phi^+_N|\rho(t)|\Phi^+_N\rangle$ and (b) Success probabilities of generating $|\Phi^+_N\rangle$ as functions of the measurement number $M$. The average excitation number of the initial coherent states for two magnon modes are set as $\beta_n=\beta_m=1,1.2,1.3$ for $N=1,2,3$, respectively. The other parameters are the same as Fig.~\ref{Fig:Coherent}.}\label{Fig:NBellStates}
\end{figure}

With proper coherent states, our scheme could be generalized to prepare a multi-excitation Bell state $|\Phi^+_N\rangle=(|00\rangle+|NN\rangle)/\sqrt{2}$ that is encoded in the ground state and a high-Fock state of the magnon modes. To hold the populations on both $|00\rangle$ and $|NN\rangle$, the measurement interval $\tau$ could be chosen such that $|\alpha_{NN}(\tau)|=1$ with $\tau=2\pi/\Omega_{NN}$ and the Rabi frequency $\Omega_{NN}=(G_e^2N+G_f^2N+\Delta^2/4)^{1/2}$. Following a similar derivation as from Eq.~(\ref{Vgtau0}) through Eq.~(\ref{EffectiveParityOperator}), one can obtain an effective projection operator $|00\rangle\langle00|+|NN\rangle\langle NN|$. In Fig.~\ref{Fig:NBellStates}(a) and Fig.~\ref{Fig:NBellStates}(b), we use various $N$ to evaluate our scheme in terms of state fidelity and success probability, respectively. A multi-excitation Bell state could be generated. The scheme becomes inefficient for a larger $N$, which results from a more dispersive distribution for the populations over a larger number of undesired Fock states between $|00\rangle$ and $|NN\rangle$. For $N=2$, the fidelity is $F=0.96$ when $M=50$ and enhanced to $F=0.98$ when $M=100$. For $N=3$, $F=0.68$ when $M=100$ and $F=0.98$ when $M=10^3$. Again, Fig.~\ref{Fig:NBellStates}(b) supports that the initial population over the target state serves as a lower bound for the success probability. The success probability declines as $N$ increases. Their final values for $N=1,2,3$ are $P_s=0.28,0.12,0.08$, respectively, which are consistent with the initial fidelities shown in Fig.~\ref{Fig:NBellStates}(a). If the magnon modes could be prepared in a superposed state of high-Fock basis $(|0\rangle+|N\rangle)\otimes(|0\rangle+|N\rangle)/2$, then the number of measurements could be much reduced as in Fig.~\ref{Fig:BellFidelity}.

\section{Single-shot scheme}\label{Sec:SingleShot}

\begin{figure}[htbp]
\centering
\includegraphics[width=0.9\linewidth]{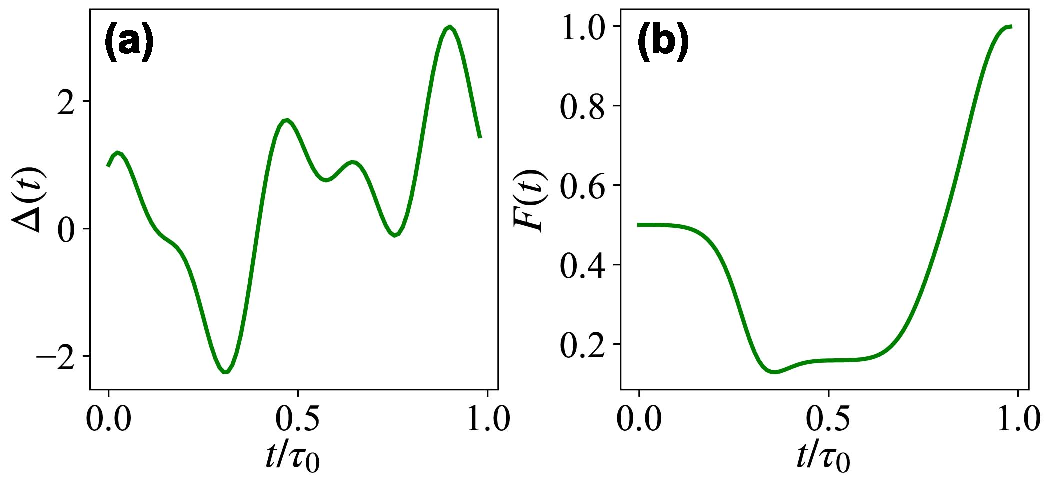}
\caption{(a) Optimized time-dependent detuning $\Delta(t)$ found with $N_\omega=4$ and a total measurement interval $\tau=\tau_0$. (b) Bell-state fidelity evolution determined by $\Delta(t)$ in (a). The initial state is the same as Fig.~(\ref{Fig:BellFidelity}). }\label{Fig:SingleShot}
\end{figure}

According to Eqs.~(\ref{ParityOperator}) and (\ref{EffectiveParityOperator}) with a fixed $\Delta$, constructing an effective parity-measurement operator demands more than one projective measurement on the ground state of the ancillary qutrit, that would enhance the overhead in experiments and is under the constraint of the lifetime of quantum system. To obtain a single-shot scheme with $M=1$ for generating the Bell state, we can manipulate the detunings in Hamiltonian~(\ref{Ham}) following a similar pattern in Ref.~\cite{CoolingNonlinear} during a period of free evolution and then perform merely one measurement on the ground state of the qutrit. For simplicity, the two detunings in Eq.~(\ref{Ham}) are assumed to have the same magnitude $G_e=G_f=G$ and are tunable in time domain. Then the full time-dependent Hamiltonian becomes
\begin{equation}
H_{\rm eff}(t)=\Delta(t)(|e\rangle\langle e|+|f\rangle\langle f|)+G(n\sigma_{eg}^++m\sigma_{fg}^++{\rm H.c.}).
\end{equation}
Aiming for the target state $|\Phi^+\rangle$, the function $\Delta(t)$ could be designed using the chopped-random basis approximation (CRAB)~\cite{CRABpra,CRABprl} and the Nelder-Mead search algorithm~\cite{NelderMead}. We take the single-round evolution period $\tau_0$ in Eq.~(\ref{Vgtau0}) as the total control time and fix the boundary condition $\Delta(0)/G=\Delta(\tau_0)/G=1$. Then the task for optimizing the time-dependent detuning is equivalent to finding an optimal combination of coefficients $a_n$ and $b_n$ in
\begin{equation}
\Delta(t)/G=1+t(\tau_0-t)\sum_{n=1}^{N_\omega}[a_n\cos(\omega_nt)+b_n\sin(\omega_nt)],
\end{equation}
where $\omega_n=2\pi n/\tau_0$. Figure~\ref{Fig:SingleShot}(a) demonstrates the common detuning $\Delta$ as a function of time obtained by the CRAB optimization with $N_\omega=4$; and Fig.~\ref{Fig:SingleShot}(b) provides the time evolution of fidelity with respect to the Bell state determined by $\Delta(t)$ in Fig.~\ref{Fig:SingleShot}(a). It is found that a magnon Bell state with a fidelity over $F=0.998$ via a near-to-perfect parity measurement of $V_g(\tau_0)=|00\rangle\langle00|+|11\rangle\langle11|$ could be realized under such an optimized Hamiltonian engineering. It indicates that we are able to generate a macroscopic entangled state with a single-shot measurement.

\section{Conclusion}\label{Sec:Conclusion}

In summary, we proposed an entangled-state generation scheme based on the parity measurement over two magnon modes, which is induced by the repeated projective measurements on the ground state of the ancillary qutrit. We demonstrate that our scheme can be practiced in a hybrid magonic system, where the dispersive interaction between magnon modes and superconducting qutrit are induced by microwave photon-magnon coupling and photon-qutrit coupling. Our scheme is dramatically distinct from those based on nonlinear interaction or squeezing Hamiltonian and the Bell state can be generated from arbitrary separable state that has a nonvanishing population in the subspace with the desired parity. The target entangled state is insensitive to the measurement imperfection and can be stabilized by our projective measurements against the environmental decoherence. We also propose a single measurement version for our scheme. Our work offers accessibility to generate Bell states in a macroscopic quantum system. In a broad perspective, it enriches the quantum control based on quantum measurement and distinguishes the efficiency of non-Gaussian operations~\cite{CatStateHe,BECSuperposition,SuperpositionHe} on non-Gaussian state generation.

\section*{Acknowledgments}

We thank Da Xu and Xu-Ke Gu for discussions about the experimental realization of the indirect coupling between magnon and superconducting qutrit. We acknowledge financial support from the National Natural Science Foundation of China (Grant No. 11974311).

\appendix
\section{Magnon-qutrit coupling mediated by a single cavity mode}\label{AppendSec:OneModeCoupling}

\begin{figure}[htbp]
\centering
\includegraphics[width=0.8\linewidth]{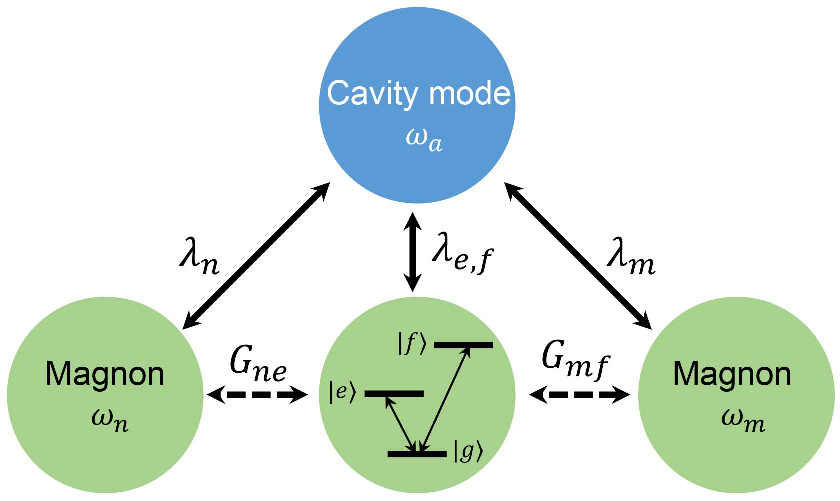}
\caption{Two magnon modes are coupled to a common cavity mode-$a$ with coupling strengths $\lambda_m$ and $\lambda_n$, respectively. The two transitions $|g\rangle\leftrightarrow|e\rangle$ and $|g\rangle\leftrightarrow|f\rangle$ in a $V$-type three-level system are coupled to cavity mode with coupling strengths $\lambda_e$ and $\lambda_f$, respectively. As an alternative model with respect to Fig.~\ref{Fig:Model}(c), it also yields the same effective Hamiltonian~(\ref{Ham}) in the dispersive regime. }\label{Fig:Model_singlemode}
\end{figure}

The model in Fig.~\ref{Fig:Model}(c) can be simplified to a scenario in Fig.~\ref{Fig:Model_singlemode} with a common cavity mode, which couples to the two transitions of qutrit in the same time. In this case, the full Hamiltonian becomes
\begin{equation}
\begin{aligned}
H=&\omega_aa^\dagger a+\omega_n n^\dagger n+\omega_m m^\dagger m +\omega_e |e\rangle\langle e|+\omega_f |f\rangle\langle f|\\ +&\lambda_n(a^\dagger n + an^\dagger) +\lambda_m(a^\dagger m + am^\dagger)\\
+& \lambda_e(a^\dagger\sigma_{eg}^-+a\sigma_{eg}^+)+ \lambda_f(a^\dagger\sigma_{fg}^-+a\sigma_{fg}^+),
\end{aligned}
\end{equation}
where $\lambda_{n,m}$ are the coupling strengths between the cavity mode and magnon modes and $\lambda_{e,f}$ are the coupling strengths between the cavity mode and the qutrit transitions. In the rotating frame with respect to
\begin{equation}
\begin{aligned}
S=&\frac{\lambda_n}{\Delta_n}(an^\dagger-a^\dagger n)+\frac{\lambda_m}{\Delta_m}(am^\dagger - a^\dagger m)\\
&+\frac{\lambda_e}{\Delta_e}(a\sigma_{eg}^+ - a^\dagger \sigma_{eg}^-)+\frac{\lambda_f}{\Delta_f}(a\sigma_{fg}^+-a^\dagger\sigma_{fg}^-)
\end{aligned}
\end{equation}
with $\Delta_i\equiv\omega_i-\omega_a$, $i=n,m,e,f$, the effective Hamiltonian to the second order of $\lambda_i/\Delta_i$ reads
\begin{equation}\label{HeffCompletesinglemode}
\begin{aligned}
\tilde{H}&=\tilde{\omega}_aa^\dagger a+\tilde{\omega}_nn^\dagger n+\tilde{\omega}_m m^\dagger m+\tilde{\omega}_e|e\rangle\langle e|+\tilde{\omega}_f|f\rangle\langle f|\\
&+G_{ne}(n\sigma_{eg}^+ + n^\dagger\sigma_{eg}^-) + G_{nf}(n\sigma_{fg}^+ + n^\dagger\sigma_{fg}^-)\\
&+G_{me}(m\sigma_{eg}^+ + m^\dagger\sigma_{eg}^-) + G_{mf}(m\sigma_{fg}^+ + m^\dagger\sigma_{fg}^-)\\
&+G_{nm}(nm^\dagger + n^\dagger m) + G_{fe}a^\dagger a\sigma_{fe}^x \\
&+ a^\dagger a(\chi_e\sigma_{eg}^z + \chi_f\sigma_{fg}^z),
\end{aligned}
\end{equation}
where
\begin{equation}
\begin{aligned}
&\tilde{\omega}_a=\omega_a-\chi_n-\chi_m,\quad\tilde{\omega}_n=\omega_n+\chi_n,\\
&\tilde{\omega}_m=\omega_m+\chi_m,\quad\tilde{\omega}_e=\omega_e+\chi_e,\quad\tilde{\omega}_f=\omega_f+\chi_f
\end{aligned}
\end{equation}
with the Lamb shifts $\chi_i=\lambda_i^2/\Delta_i$, $\sigma_{ij}^x\equiv|i\rangle\langle j|+|j\rangle\langle i|$ and $\sigma_{ij}^z\equiv|i\rangle\langle i|-|j\rangle\langle j|$. The coupling strengths induced by the cavity mode could be expressed as
\begin{equation}
G_{ij}=\frac{\lambda_i\lambda_j}{2}\left(\frac{1}{\Delta_i}+\frac{1}{\Delta_j}\right).
\end{equation}
In comparison to the dispersively induced effective Hamiltonian~(\ref{HeffComplete}) for the two-cavity-mode situation, here the cross interactions emerge in the effective Hamiltonian~(\ref{HeffCompletesinglemode}) for the single-cavity-mode situation. They include the interaction $G_{nf}$ between magnon-$n$ and transition $|g\rangle\leftrightarrow|f\rangle$, the interaction $G_{me}$ between magnon-$m$ and transition $|g\rangle\leftrightarrow|e\rangle$, and the interaction $G_{fe}$ between two excited levels in qutrit. However, these terms could be wisely neutralized under the detuning-match condition:
\begin{equation}\label{DetuningCondition}
 \Delta_n=\Delta_e=-\Delta_m=-\Delta_f.
\end{equation}
Together with the vacuum-state assumption $\langle a^\dagger a\rangle\approx0$, the effective Hamiltonian becomes
\begin{equation}
\begin{aligned}
\tilde{H}&\approx\tilde{\omega}_aa^\dagger a+\tilde{\omega}_nn^\dagger n+\tilde{\omega}_m m^\dagger m+\tilde{\omega}_e|e\rangle\langle e|+\tilde{\omega}_f|f\rangle\langle f|\\
&+ G_{ne}(n\sigma_{eg}^++n^\dagger\sigma_{eg}^-) + G_{mf}(m\sigma_{fg}^++m^\dagger\sigma_{fg}^-).
\end{aligned}
\end{equation}
In the rotating frame with respect to $H_R=\tilde{\omega}_aa^\dagger a + \tilde{\omega}_n(n^\dagger n+|e\rangle\langle e|)+\tilde{\omega}_m(m^\dagger m+|f\rangle\langle f|)$, we have exactly the same form as Hamiltonian in Eq.~(\ref{Ham}):
\begin{equation}
\begin{aligned}
H_{\rm eff}=&\tilde{\Delta}_e|e\rangle\langle e|+\tilde{\Delta}_f|f\rangle\langle f|+G_{ne}(n\sigma_{eg}^++n^\dagger\sigma_{eg}^-)\\
 &+ G_{mf}(m\sigma_{fg}^++m^\dagger\sigma_{fg}^-)
\end{aligned}
\end{equation}
with $\tilde{\Delta}_e=\tilde{\omega}_e-\tilde{\omega}_n$ and $\tilde{\Delta}_f=\tilde{\omega}_f-\tilde{\omega}_m$.

\section{Effective measurement operator with detection noises}\label{AppendSec:OperatorWithNoise}

With detection noise in each projective measurement, the measurement operator can be modified from $M_g=|g\rangle\langle g|$ to $M_{\tilde{g}}=|\tilde{g}\rangle\langle\tilde{g}|$, where $|\tilde{g}\rangle=|g\rangle+\epsilon_e|e\rangle+\epsilon_f|f\rangle$. The deviation ratios $\epsilon_e$ and $\epsilon_f$ are assumed to be the same magnitude $\epsilon$ for simplicity. To the first order of $\epsilon$, the nonunitary evolution operator after one round of free evolution and measurement can be written as
\begin{equation}\label{VoneMeasurement}
V_{\tilde{g}}=\langle\tilde{g}|U(\tau)|\tilde{g}\rangle\approx V_g(\tau)+\epsilon\mathcal{W}(\tau),
\end{equation}
where $V_g(\tau)$ is the nonunitary evolution operator in Eq.~(\ref{KrausOperator}) and
\begin{equation}
\mathcal{W}(\tau)=\langle{g}|U(\tau)|e\rangle+\langle e|U(\tau)|g\rangle+\langle f|U(\tau)|g\rangle+\langle g|U(\tau)|f\rangle
\end{equation}
results from the imperfect measurement and yields the unwanted off-diagonal transitions. In particular, we have
\begin{equation}
\begin{aligned}
&\langle g|U(\tau)|e\rangle=e^{-i\Delta\tau/2}\sum_{n,m}\mu_{nm}(\tau)|nm\rangle\langle n-1,m|,\\
&\langle e|U(\tau)|g\rangle=e^{-i\Delta\tau/2}\sum_{n,m}\mu_{nm}(\tau)|n-1,m\rangle\langle nm|, \\
&\langle g|U(\tau)|f\rangle=e^{-i\Delta\tau/2}\sum_{n,m}\nu_{nm}(\tau)|nm\rangle\langle n,m-1|,\\
&\langle f|U(\tau)|g\rangle=e^{-i\Delta\tau/2}\sum_{n,m}\nu_{nm}(\tau)|n,m-1\rangle\langle nm|,
\end{aligned}
\end{equation}
where
\begin{equation}
\begin{aligned}
&\mu_{nm}(\tau)\equiv-i\frac{G_e\sqrt{n}\sin(\Omega_{nm}\tau)}{\Omega_{nm}},\\
&\nu_{nm}(\tau)\equiv-i\frac{G_f\sqrt{m}\sin(\Omega_{nm}\tau)}{\Omega_{nm}}.
\end{aligned}
\end{equation}
After $M$ measurements, the nonunitary evolution operator (to the first order of $\epsilon$) becomes
\begin{equation}
\begin{aligned}
V_{\tilde{g}}^M&\approx V_g^M(\tau)+\epsilon MV_g^{M-1}(\tau)\mathcal{W}(\tau) \\
&\approx\alpha_{00}^M|00\rangle\langle00|+\alpha_{11}^M|11\rangle\langle 11|+\alpha_{10}^M|10\rangle\langle 10|+\alpha_{01}^M|01\rangle\langle 01|\\ &+\epsilon M\bigg[\alpha_{10}^{M-1}\mu_{10}|10\rangle\langle00|+\alpha_{00}^{M-1}\mu_{10}|00\rangle\langle10|
\\
&+\alpha_{01}^{M-1}\nu_{01}|01\rangle\langle00|+\alpha_{00}^{M-1}\nu_{01}|00\rangle\langle01|\\
&+\alpha_{11}^{M-1}\mu_{11}|11\rangle\langle01|+\alpha_{01}^{M-1}\mu_{11}|01\rangle\langle11|\\
&+\alpha_{11}^{M-1}\nu_{11}|11\rangle\langle10|+\alpha_{10}^{M-1}\nu_{11}|10\rangle\langle11|\bigg]
\end{aligned}
\end{equation}
under the assumptions that the two magnon modes are near-resonant to the relevant qutrit transitions and none of them is double excited. Note $\alpha_{00}=1$ is always valid and $|\alpha_{11}(\tau_0)|=1$ is valid under a properly chosen interval $\tau=\tau_0$. The rest terms, such as $\alpha_{10}^M$ and $\alpha_{01}^M$, will vanish in an exponential way since $|\alpha_{10}|,|\alpha_{01}|<1$. Then eventually the effective measurement operator with detection noises reads
\begin{equation}
\begin{aligned}
\tilde{V}_{\tilde{g}}^M&\approx|00\rangle\langle00|+|11\rangle\langle 11|+\epsilon M\bigg[\mu_{10}|00\rangle\langle10|+\nu_{01}|00\rangle\langle01|\\
&+\mu_{11}|11\rangle\langle01|+\nu_{11}|11\rangle\langle10|\bigg].
\end{aligned}
\end{equation}

\section{Coupling-ratio optimization in Bell state generation}\label{AppendSec:OptimalCoupling}

\begin{figure}[htbp]
\centering
\includegraphics[width=0.9\linewidth]{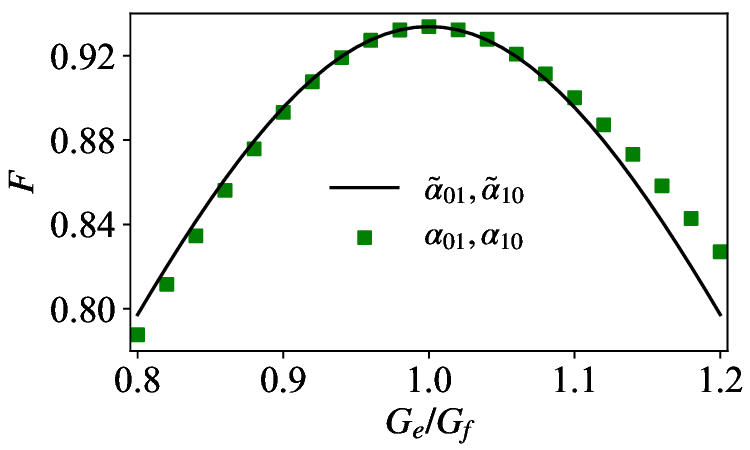}
\caption{Fidelities of the Bell state $|\Phi^+\rangle$ under a single parity measurement as a function of the coupling-strength ratio $G_e/G_f$ with and without the second-order approximation in Eq.~(\ref{CoeffWithApprox}). The other parameters are the same as those in Fig.~\ref{Fig:BellFidelity}. }\label{Fig:Coupling}
\end{figure}

This appendix contributes to estimating the effect of the ratio $G_e/G_f$ about magnons-qutrit coupling strengths on Bell state generation. For simplicity, we consider the same initial condition in Fig.~\ref{Fig:BellFidelity}. And in the near-resonant regime, the measurement-induced evolution operator for the two magnon modes in Eq.~(\ref{Vgtau0}) can be written as
\begin{equation}
V_g(\tau_0)\approx|00\rangle\langle00|+|11\rangle\langle11|+\alpha_{01}|01\rangle\langle01|+\alpha_{10}|10\rangle\langle10|
\end{equation}
with
\begin{equation}\label{Coeff}
\alpha_{01}=\cos\left(\frac{2\pi G_f}{\sqrt{G_e^2+G_f^2}}\right),\quad \alpha_{10}=\cos\left(\frac{2\pi G_e}{\sqrt{G_e^2+G_f^2}}\right),
\end{equation}
where the measurement interval is fixed as $\tau=\tau_0=2\pi/(G_e^2+G_f^2)^{1/2}$. Then after a single round of evolution and measurement, the fidelity of the Bell state can be written as
\begin{equation}\label{Falpha}
F=\frac{|\langle\Phi^+|V_g(\tau)|\psi_i\rangle|^2}{{\rm Tr}[V_g(\tau)|\psi_i\rangle\langle\psi_i|V_g^\dagger(\tau)]}=\frac{2}{2+\alpha_{01}^2(\tau)+\alpha_{10}^2(\tau)}.
\end{equation}
Accordingly, the nonvanishing coefficients $\alpha_{10}(\tau)$ and $\alpha_{01}(\tau)$ result in an imperfect parity measurement by reducing the final Bell-state fidelity. When $G_e\approx G_f$, both coefficients can be expanded around the coupling ratio $\xi\equiv G_f/G_e=1$. To the order of $\mathcal{O}[(\xi-1)^2]$, we have
\begin{equation}\label{CoeffWithApprox}
\begin{aligned}
&\alpha_{01}(\tau)\approx\tilde{\alpha}_{01}(\tau)=\cos(\sqrt{2}\pi)-\frac{\pi}{\sqrt{2}}\sin(\sqrt{2}\pi)(\xi-1)\\
&\alpha_{10}(\tau)\approx\tilde{\alpha}_{10}(\tau)=\cos(\sqrt{2}\pi)+\frac{\pi}{\sqrt{2}}\sin(\sqrt{2}\pi)(\xi-1).
\end{aligned}
\end{equation}
Then the denominator in Eq.~(\ref{Falpha}) depends on
\begin{equation}
\alpha_{01}^2+\alpha_{10}^2=\cos^2(\sqrt{2}\pi)+\frac{\pi^2}{2}\sin^2(\sqrt{2}\pi)(\xi-1)^2.
\end{equation}
It is straightforward to see that the Bell-state generation is optimized when $\xi=1$, i.e., $G_e/G_f=1$. In Fig.~\ref{Fig:Coupling}, the Bell-state fidelities are plotted with the coefficients in Eq.~(\ref{Coeff}) and those in Eq.~(\ref{CoeffWithApprox}). It is shown that Eq.~(\ref{CoeffWithApprox}) is a good approximation to Eq.~(\ref{Coeff}). And the numerical simulations confirm the coupling-ratio optimization condition.

\section*{References}
\bibliography{ref_APL}

\begin{thebibliography}{60}%
\makeatletter
\providecommand \@ifxundefined [1]{%
 \@ifx{#1\undefined}
}%
\providecommand \@ifnum [1]{%
 \ifnum #1\expandafter \@firstoftwo
 \else \expandafter \@secondoftwo
 \fi
}%
\providecommand \@ifx [1]{%
 \ifx #1\expandafter \@firstoftwo
 \else \expandafter \@secondoftwo
 \fi
}%
\providecommand \natexlab [1]{#1}%
\providecommand \enquote  [1]{``#1''}%
\providecommand \bibnamefont  [1]{#1}%
\providecommand \bibfnamefont [1]{#1}%
\providecommand \citenamefont [1]{#1}%
\providecommand \href@noop [0]{\@secondoftwo}%
\providecommand \href [0]{\begingroup \@sanitize@url \@href}%
\providecommand \@href[1]{\@@startlink{#1}\@@href}%
\providecommand \@@href[1]{\endgroup#1\@@endlink}%
\providecommand \@sanitize@url [0]{\catcode `\\12\catcode `\$12\catcode
  `\&12\catcode `\#12\catcode `\^12\catcode `\_12\catcode `\%12\relax}%
\providecommand \@@startlink[1]{}%
\providecommand \@@endlink[0]{}%
\providecommand \url  [0]{\begingroup\@sanitize@url \@url }%
\providecommand \@url [1]{\endgroup\@href {#1}{\urlprefix }}%
\providecommand \urlprefix  [0]{URL }%
\providecommand \Eprint [0]{\href }%
\providecommand \doibase [0]{http://dx.doi.org/}%
\providecommand \selectlanguage [0]{\@gobble}%
\providecommand \bibinfo  [0]{\@secondoftwo}%
\providecommand \bibfield  [0]{\@secondoftwo}%
\providecommand \translation [1]{[#1]}%
\providecommand \BibitemOpen [0]{}%
\providecommand \bibitemStop [0]{}%
\providecommand \bibitemNoStop [0]{.\EOS\space}%
\providecommand \EOS [0]{\spacefactor3000\relax}%
\providecommand \BibitemShut  [1]{\csname bibitem#1\endcsname}%
\let\auto@bib@innerbib\@empty
%</preamble>
\bibitem [{\citenamefont {Misra}\ and\ \citenamefont
  {Sudarshan}(1977)}]{ZenoParadox}%
  \BibitemOpen
  \bibfield  {author} {\bibinfo {author} {\bibfnamefont {B.}~\bibnamefont
  {Misra}}\ and\ \bibinfo {author} {\bibfnamefont {E.~C.~G.}\ \bibnamefont
  {Sudarshan}},\ }\bibfield  {title} {\enquote {\bibinfo {title} {The zeno’s
  paradox in quantum theory},}\ }\href {\doibase 10.1063/1.523304} {\bibfield
  {journal} {\bibinfo  {journal} {J. Math. Phys.}\ }\textbf {\bibinfo {volume}
  {18}},\ \bibinfo {pages} {756--763} (\bibinfo {year} {1977})}\BibitemShut
  {NoStop}%
\bibitem [{\citenamefont {Aharonov}\ and\ \citenamefont
  {Vardi}(1980)}]{AharonovZeno}%
  \BibitemOpen
  \bibfield  {author} {\bibinfo {author} {\bibfnamefont {Y.}~\bibnamefont
  {Aharonov}}\ and\ \bibinfo {author} {\bibfnamefont {M.}~\bibnamefont
  {Vardi}},\ }\bibfield  {title} {\enquote {\bibinfo {title} {Meaning of an
  individual ``feynman path"},}\ }\href {\doibase 10.1103/PhysRevD.21.2235}
  {\bibfield  {journal} {\bibinfo  {journal} {Phys. Rev. D}\ }\textbf {\bibinfo
  {volume} {21}},\ \bibinfo {pages} {2235--2240} (\bibinfo {year}
  {1980})}\BibitemShut {NoStop}%
\bibitem [{\citenamefont {Altenm\"uller}\ and\ \citenamefont
  {Schenzle}(1993)}]{NonselectiveAharonovZeno}%
  \BibitemOpen
  \bibfield  {author} {\bibinfo {author} {\bibfnamefont {T.~P.}\ \bibnamefont
  {Altenm\"uller}}\ and\ \bibinfo {author} {\bibfnamefont {A.}~\bibnamefont
  {Schenzle}},\ }\bibfield  {title} {\enquote {\bibinfo {title} {Dynamics by
  measurement: Aharonov's inverse quantum zeno effect},}\ }\href {\doibase
  10.1103/PhysRevA.48.70} {\bibfield  {journal} {\bibinfo  {journal} {Phys.
  Rev. A}\ }\textbf {\bibinfo {volume} {48}},\ \bibinfo {pages} {70--79}
  (\bibinfo {year} {1993})}\BibitemShut {NoStop}%
\bibitem [{\citenamefont {Roa}\ \emph {et~al.}(2006)\citenamefont {Roa},
  \citenamefont {Delgado}, \citenamefont {Ladr\'on~de Guevara},\ and\
  \citenamefont {Klimov}}]{MeasurementEvolution}%
  \BibitemOpen
  \bibfield  {author} {\bibinfo {author} {\bibfnamefont {L.}~\bibnamefont
  {Roa}}, \bibinfo {author} {\bibfnamefont {A.}~\bibnamefont {Delgado}},
  \bibinfo {author} {\bibfnamefont {M.~L.}\ \bibnamefont {Ladr\'on~de
  Guevara}}, \ and\ \bibinfo {author} {\bibfnamefont {A.~B.}\ \bibnamefont
  {Klimov}},\ }\bibfield  {title} {\enquote {\bibinfo {title}
  {Measurement-driven quantum evolution},}\ }\href {\doibase
  10.1103/PhysRevA.73.012322} {\bibfield  {journal} {\bibinfo  {journal} {Phys.
  Rev. A}\ }\textbf {\bibinfo {volume} {73}},\ \bibinfo {pages} {012322}
  (\bibinfo {year} {2006})}\BibitemShut {NoStop}%
\bibitem [{\citenamefont {Pechen}\ \emph {et~al.}(2006)\citenamefont {Pechen},
  \citenamefont {Il'in}, \citenamefont {Shuang},\ and\ \citenamefont
  {Rabitz}}]{NonselectiveMeasurementEvolution}%
  \BibitemOpen
  \bibfield  {author} {\bibinfo {author} {\bibfnamefont {A.}~\bibnamefont
  {Pechen}}, \bibinfo {author} {\bibfnamefont {N.}~\bibnamefont {Il'in}},
  \bibinfo {author} {\bibfnamefont {F.}~\bibnamefont {Shuang}}, \ and\ \bibinfo
  {author} {\bibfnamefont {H.}~\bibnamefont {Rabitz}},\ }\bibfield  {title}
  {\enquote {\bibinfo {title} {Quantum control by von neumann measurements},}\
  }\href {\doibase 10.1103/PhysRevA.74.052102} {\bibfield  {journal} {\bibinfo
  {journal} {Phys. Rev. A}\ }\textbf {\bibinfo {volume} {74}},\ \bibinfo
  {pages} {052102} (\bibinfo {year} {2006})}\BibitemShut {NoStop}%
\bibitem [{\citenamefont {Nakazato}, \citenamefont {Takazawa},\ and\
  \citenamefont {Yuasa}(2003)}]{MeasurementPurification}%
  \BibitemOpen
  \bibfield  {author} {\bibinfo {author} {\bibfnamefont {H.}~\bibnamefont
  {Nakazato}}, \bibinfo {author} {\bibfnamefont {T.}~\bibnamefont {Takazawa}},
  \ and\ \bibinfo {author} {\bibfnamefont {K.}~\bibnamefont {Yuasa}},\
  }\bibfield  {title} {\enquote {\bibinfo {title} {Purification through
  zeno-like measurements},}\ }\href {\doibase 10.1103/PhysRevLett.90.060401}
  {\bibfield  {journal} {\bibinfo  {journal} {Phys. Rev. Lett.}\ }\textbf
  {\bibinfo {volume} {90}},\ \bibinfo {pages} {060401} (\bibinfo {year}
  {2003})}\BibitemShut {NoStop}%
\bibitem [{\citenamefont {Yan}\ and\ \citenamefont
  {Jing}(2023{\natexlab{a}})}]{EntanglementPurification}%
  \BibitemOpen
  \bibfield  {author} {\bibinfo {author} {\bibfnamefont {J.-s.}\ \bibnamefont
  {Yan}}\ and\ \bibinfo {author} {\bibfnamefont {J.}~\bibnamefont {Jing}},\
  }\bibfield  {title} {\enquote {\bibinfo {title} {Generic eigenstate
  preparation via measurement-based purification},}\ }\href {\doibase
  10.1103/PhysRevA.108.042215} {\bibfield  {journal} {\bibinfo  {journal}
  {Phys. Rev. A}\ }\textbf {\bibinfo {volume} {108}},\ \bibinfo {pages}
  {042215} (\bibinfo {year} {2023}{\natexlab{a}})}\BibitemShut {NoStop}%
\bibitem [{\citenamefont {Paulsen}(2003)}]{Naimark}%
  \BibitemOpen
  \bibfield  {author} {\bibinfo {author} {\bibfnamefont {V.}~\bibnamefont
  {Paulsen}},\ }\href@noop {} {\emph {\bibinfo {title} {Completely Bounded Maps
  and Operator Algebras}}}\ (\bibinfo  {publisher} {Cambridge University
  Press},\ \bibinfo {address} {Cambridge},\ \bibinfo {year} {2003})\BibitemShut
  {NoStop}%
\bibitem [{\citenamefont {Li}\ \emph {et~al.}(2011)\citenamefont {Li},
  \citenamefont {Wu}, \citenamefont {Wang},\ and\ \citenamefont
  {Yang}}]{MeasurementCooling}%
  \BibitemOpen
  \bibfield  {author} {\bibinfo {author} {\bibfnamefont {Y.}~\bibnamefont
  {Li}}, \bibinfo {author} {\bibfnamefont {L.-A.}\ \bibnamefont {Wu}}, \bibinfo
  {author} {\bibfnamefont {Y.-D.}\ \bibnamefont {Wang}}, \ and\ \bibinfo
  {author} {\bibfnamefont {L.-P.}\ \bibnamefont {Yang}},\ }\bibfield  {title}
  {\enquote {\bibinfo {title} {Nondeterministic ultrafast ground-state cooling
  of a mechanical resonator},}\ }\href {\doibase 10.1103/PhysRevB.84.094502}
  {\bibfield  {journal} {\bibinfo  {journal} {Phys. Rev. B}\ }\textbf {\bibinfo
  {volume} {84}},\ \bibinfo {pages} {094502} (\bibinfo {year}
  {2011})}\BibitemShut {NoStop}%
\bibitem [{\citenamefont {Buffoni}\ \emph {et~al.}(2019)\citenamefont
  {Buffoni}, \citenamefont {Solfanelli}, \citenamefont {Verrucchi},
  \citenamefont {Cuccoli},\ and\ \citenamefont
  {Campisi}}]{MeasurementCoolingPRL}%
  \BibitemOpen
  \bibfield  {author} {\bibinfo {author} {\bibfnamefont {L.}~\bibnamefont
  {Buffoni}}, \bibinfo {author} {\bibfnamefont {A.}~\bibnamefont {Solfanelli}},
  \bibinfo {author} {\bibfnamefont {P.}~\bibnamefont {Verrucchi}}, \bibinfo
  {author} {\bibfnamefont {A.}~\bibnamefont {Cuccoli}}, \ and\ \bibinfo
  {author} {\bibfnamefont {M.}~\bibnamefont {Campisi}},\ }\bibfield  {title}
  {\enquote {\bibinfo {title} {Quantum measurement cooling},}\ }\href {\doibase
  10.1103/PhysRevLett.122.070603} {\bibfield  {journal} {\bibinfo  {journal}
  {Phys. Rev. Lett.}\ }\textbf {\bibinfo {volume} {122}},\ \bibinfo {pages}
  {070603} (\bibinfo {year} {2019})}\BibitemShut {NoStop}%
\bibitem [{\citenamefont {Yan}\ and\ \citenamefont
  {Jing}(2021)}]{ExternalLevelCooling}%
  \BibitemOpen
  \bibfield  {author} {\bibinfo {author} {\bibfnamefont {J.-s.}\ \bibnamefont
  {Yan}}\ and\ \bibinfo {author} {\bibfnamefont {J.}~\bibnamefont {Jing}},\
  }\bibfield  {title} {\enquote {\bibinfo {title} {External-level assisted
  cooling by measurement},}\ }\href {\doibase 10.1103/PhysRevA.104.063105}
  {\bibfield  {journal} {\bibinfo  {journal} {Phys. Rev. A}\ }\textbf {\bibinfo
  {volume} {104}},\ \bibinfo {pages} {063105} (\bibinfo {year}
  {2021})}\BibitemShut {NoStop}%
\bibitem [{\citenamefont {Xu}\ \emph {et~al.}(2014)\citenamefont {Xu},
  \citenamefont {Yung}, \citenamefont {Xu}, \citenamefont {Boixo},
  \citenamefont {Zhou}, \citenamefont {Li}, \citenamefont {Aspuru-Guzik},\ and\
  \citenamefont {Guo}}]{ExpOneModeCooling}%
  \BibitemOpen
  \bibfield  {author} {\bibinfo {author} {\bibfnamefont {J.-S.}\ \bibnamefont
  {Xu}}, \bibinfo {author} {\bibfnamefont {M.-H.}\ \bibnamefont {Yung}},
  \bibinfo {author} {\bibfnamefont {X.-Y.}\ \bibnamefont {Xu}}, \bibinfo
  {author} {\bibfnamefont {S.}~\bibnamefont {Boixo}}, \bibinfo {author}
  {\bibfnamefont {Z.-W.}\ \bibnamefont {Zhou}}, \bibinfo {author}
  {\bibfnamefont {C.-F.}\ \bibnamefont {Li}}, \bibinfo {author} {\bibfnamefont
  {A.}~\bibnamefont {Aspuru-Guzik}}, \ and\ \bibinfo {author} {\bibfnamefont
  {G.-C.}\ \bibnamefont {Guo}},\ }\bibfield  {title} {\enquote {\bibinfo
  {title} {Demon-like algorithmic quantum cooling and its realization with
  quantum optics},}\ }\href {\doibase 10.1038/nphoton.2013.354} {\bibfield
  {journal} {\bibinfo  {journal} {Nat. Photonics}\ }\textbf {\bibinfo {volume}
  {8}},\ \bibinfo {pages} {113--118} (\bibinfo {year} {2014})}\BibitemShut
  {NoStop}%
\bibitem [{\citenamefont {Rao}, \citenamefont {Momenzadeh},\ and\ \citenamefont
  {Wrachtrup}(2016)}]{HeraldedControlMotion}%
  \BibitemOpen
  \bibfield  {author} {\bibinfo {author} {\bibfnamefont {D.~D.~B.}\
  \bibnamefont {Rao}}, \bibinfo {author} {\bibfnamefont {S.~A.}\ \bibnamefont
  {Momenzadeh}}, \ and\ \bibinfo {author} {\bibfnamefont {J.}~\bibnamefont
  {Wrachtrup}},\ }\bibfield  {title} {\enquote {\bibinfo {title} {Heralded
  control of mechanical motion by single spins},}\ }\href {\doibase
  10.1103/PhysRevLett.117.077203} {\bibfield  {journal} {\bibinfo  {journal}
  {Phys. Rev. Lett.}\ }\textbf {\bibinfo {volume} {117}},\ \bibinfo {pages}
  {077203} (\bibinfo {year} {2016})}\BibitemShut {NoStop}%
\bibitem [{\citenamefont {Seah}\ \emph {et~al.}(2021)\citenamefont {Seah},
  \citenamefont {Perarnau-Llobet}, \citenamefont {Haack}, \citenamefont
  {Brunner},\ and\ \citenamefont {Nimmrichter}}]{NonselevtiveCharging}%
  \BibitemOpen
  \bibfield  {author} {\bibinfo {author} {\bibfnamefont {S.}~\bibnamefont
  {Seah}}, \bibinfo {author} {\bibfnamefont {M.}~\bibnamefont
  {Perarnau-Llobet}}, \bibinfo {author} {\bibfnamefont {G.}~\bibnamefont
  {Haack}}, \bibinfo {author} {\bibfnamefont {N.}~\bibnamefont {Brunner}}, \
  and\ \bibinfo {author} {\bibfnamefont {S.}~\bibnamefont {Nimmrichter}},\
  }\bibfield  {title} {\enquote {\bibinfo {title} {Quantum speed-up in
  collisional battery charging},}\ }\href {\doibase
  10.1103/PhysRevLett.127.100601} {\bibfield  {journal} {\bibinfo  {journal}
  {Phys. Rev. Lett.}\ }\textbf {\bibinfo {volume} {127}},\ \bibinfo {pages}
  {100601} (\bibinfo {year} {2021})}\BibitemShut {NoStop}%
\bibitem [{\citenamefont {Yan}\ and\ \citenamefont
  {Jing}(2023{\natexlab{b}})}]{MeasurementCharging}%
  \BibitemOpen
  \bibfield  {author} {\bibinfo {author} {\bibfnamefont {J.-s.}\ \bibnamefont
  {Yan}}\ and\ \bibinfo {author} {\bibfnamefont {J.}~\bibnamefont {Jing}},\
  }\bibfield  {title} {\enquote {\bibinfo {title} {Charging by quantum
  measurement},}\ }\href {\doibase 10.1103/PhysRevApplied.19.064069} {\bibfield
   {journal} {\bibinfo  {journal} {Phys. Rev. Appl.}\ }\textbf {\bibinfo
  {volume} {19}},\ \bibinfo {pages} {064069} (\bibinfo {year}
  {2023}{\natexlab{b}})}\BibitemShut {NoStop}%
\bibitem [{\citenamefont {Terhal}(2015)}]{QuantumErrorCorrection}%
  \BibitemOpen
  \bibfield  {author} {\bibinfo {author} {\bibfnamefont {B.~M.}\ \bibnamefont
  {Terhal}},\ }\bibfield  {title} {\enquote {\bibinfo {title} {Quantum error
  correction for quantum memories},}\ }\href {\doibase
  10.1103/RevModPhys.87.307} {\bibfield  {journal} {\bibinfo  {journal} {Rev.
  Mod. Phys.}\ }\textbf {\bibinfo {volume} {87}},\ \bibinfo {pages} {307--346}
  (\bibinfo {year} {2015})}\BibitemShut {NoStop}%
\bibitem [{\citenamefont {Gottesman}, \citenamefont {Kitaev},\ and\
  \citenamefont {Preskill}(2001)}]{QubitInOsillator}%
  \BibitemOpen
  \bibfield  {author} {\bibinfo {author} {\bibfnamefont {D.}~\bibnamefont
  {Gottesman}}, \bibinfo {author} {\bibfnamefont {A.}~\bibnamefont {Kitaev}}, \
  and\ \bibinfo {author} {\bibfnamefont {J.}~\bibnamefont {Preskill}},\
  }\bibfield  {title} {\enquote {\bibinfo {title} {Encoding a qubit in an
  oscillator},}\ }\href {\doibase 10.1103/PhysRevA.64.012310} {\bibfield
  {journal} {\bibinfo  {journal} {Phys. Rev. A}\ }\textbf {\bibinfo {volume}
  {64}},\ \bibinfo {pages} {012310} (\bibinfo {year} {2001})}\BibitemShut
  {NoStop}%
\bibitem [{\citenamefont {Saira}\ \emph {et~al.}(2014)\citenamefont {Saira},
  \citenamefont {Groen}, \citenamefont {Cramer}, \citenamefont {Meretska},
  \citenamefont {de~Lange},\ and\ \citenamefont
  {DiCarlo}}]{ParityMeasurementCQED}%
  \BibitemOpen
  \bibfield  {author} {\bibinfo {author} {\bibfnamefont {O.-P.}\ \bibnamefont
  {Saira}}, \bibinfo {author} {\bibfnamefont {J.~P.}\ \bibnamefont {Groen}},
  \bibinfo {author} {\bibfnamefont {J.}~\bibnamefont {Cramer}}, \bibinfo
  {author} {\bibfnamefont {M.}~\bibnamefont {Meretska}}, \bibinfo {author}
  {\bibfnamefont {G.}~\bibnamefont {de~Lange}}, \ and\ \bibinfo {author}
  {\bibfnamefont {L.}~\bibnamefont {DiCarlo}},\ }\bibfield  {title} {\enquote
  {\bibinfo {title} {Entanglement genesis by ancilla-based parity measurement
  in 2d circuit qed},}\ }\href {\doibase 10.1103/PhysRevLett.112.070502}
  {\bibfield  {journal} {\bibinfo  {journal} {Phys. Rev. Lett.}\ }\textbf
  {\bibinfo {volume} {112}},\ \bibinfo {pages} {070502} (\bibinfo {year}
  {2014})}\BibitemShut {NoStop}%
\bibitem [{\citenamefont {Rist{\`e}}\ \emph {et~al.}(2013)\citenamefont
  {Rist{\`e}}, \citenamefont {Dukalski}, \citenamefont {Watson}, \citenamefont
  {de~Lange}, \citenamefont {Tiggelman}, \citenamefont {Blanter}, \citenamefont
  {Lehnert}, \citenamefont {Schouten},\ and\ \citenamefont
  {DiCarlo}}]{EntanglementByParityMeasurement}%
  \BibitemOpen
  \bibfield  {author} {\bibinfo {author} {\bibfnamefont {D.}~\bibnamefont
  {Rist{\`e}}}, \bibinfo {author} {\bibfnamefont {M.}~\bibnamefont {Dukalski}},
  \bibinfo {author} {\bibfnamefont {C.~A.}\ \bibnamefont {Watson}}, \bibinfo
  {author} {\bibfnamefont {G.}~\bibnamefont {de~Lange}}, \bibinfo {author}
  {\bibfnamefont {M.~J.}\ \bibnamefont {Tiggelman}}, \bibinfo {author}
  {\bibfnamefont {Y.~M.}\ \bibnamefont {Blanter}}, \bibinfo {author}
  {\bibfnamefont {K.~W.}\ \bibnamefont {Lehnert}}, \bibinfo {author}
  {\bibfnamefont {R.~N.}\ \bibnamefont {Schouten}}, \ and\ \bibinfo {author}
  {\bibfnamefont {L.}~\bibnamefont {DiCarlo}},\ }\bibfield  {title} {\enquote
  {\bibinfo {title} {Deterministic entanglement of superconducting qubits by
  parity measurement and feedback},}\ }\href {\doibase 10.1038/nature12513}
  {\bibfield  {journal} {\bibinfo  {journal} {Nature}\ }\textbf {\bibinfo
  {volume} {502}},\ \bibinfo {pages} {350--354} (\bibinfo {year}
  {2013})}\BibitemShut {NoStop}%
\bibitem [{\citenamefont {Andersen}\ \emph {et~al.}(2019)\citenamefont
  {Andersen}, \citenamefont {Remm}, \citenamefont {Lazar}, \citenamefont
  {Krinner}, \citenamefont {Heinsoo}, \citenamefont {Besse}, \citenamefont
  {Gabureac}, \citenamefont {Wallraff},\ and\ \citenamefont
  {Eichler}}]{BellnpjQuanInf}%
  \BibitemOpen
  \bibfield  {author} {\bibinfo {author} {\bibfnamefont {C.~K.}\ \bibnamefont
  {Andersen}}, \bibinfo {author} {\bibfnamefont {A.}~\bibnamefont {Remm}},
  \bibinfo {author} {\bibfnamefont {S.}~\bibnamefont {Lazar}}, \bibinfo
  {author} {\bibfnamefont {S.}~\bibnamefont {Krinner}}, \bibinfo {author}
  {\bibfnamefont {J.}~\bibnamefont {Heinsoo}}, \bibinfo {author} {\bibfnamefont
  {J.-C.}\ \bibnamefont {Besse}}, \bibinfo {author} {\bibfnamefont
  {M.}~\bibnamefont {Gabureac}}, \bibinfo {author} {\bibfnamefont
  {A.}~\bibnamefont {Wallraff}}, \ and\ \bibinfo {author} {\bibfnamefont
  {C.}~\bibnamefont {Eichler}},\ }\bibfield  {title} {\enquote {\bibinfo
  {title} {Entanglement stabilization using ancilla-based parity detection and
  real-time feedback in superconducting circuits},}\ }\href {\doibase
  10.1038/s41534-019-0185-4} {\bibfield  {journal} {\bibinfo  {journal} {npj
  Quantum Inf.}\ }\textbf {\bibinfo {volume} {5}},\ \bibinfo {pages} {69}
  (\bibinfo {year} {2019})}\BibitemShut {NoStop}%
\bibitem [{\citenamefont {Blumenthal}\ \emph {et~al.}(2022)\citenamefont
  {Blumenthal}, \citenamefont {Mor}, \citenamefont {Diringer}, \citenamefont
  {Martin}, \citenamefont {Lewalle}, \citenamefont {Burgarth}, \citenamefont
  {Whaley},\ and\ \citenamefont {Hacohen-Gourgy}}]{ZenoGate}%
  \BibitemOpen
  \bibfield  {author} {\bibinfo {author} {\bibfnamefont {E.}~\bibnamefont
  {Blumenthal}}, \bibinfo {author} {\bibfnamefont {C.}~\bibnamefont {Mor}},
  \bibinfo {author} {\bibfnamefont {A.~A.}\ \bibnamefont {Diringer}}, \bibinfo
  {author} {\bibfnamefont {L.~S.}\ \bibnamefont {Martin}}, \bibinfo {author}
  {\bibfnamefont {P.}~\bibnamefont {Lewalle}}, \bibinfo {author} {\bibfnamefont
  {D.}~\bibnamefont {Burgarth}}, \bibinfo {author} {\bibfnamefont {K.~B.}\
  \bibnamefont {Whaley}}, \ and\ \bibinfo {author} {\bibfnamefont
  {S.}~\bibnamefont {Hacohen-Gourgy}},\ }\bibfield  {title} {\enquote {\bibinfo
  {title} {Demonstration of universal control between non-interacting qubits
  using the quantum zeno effect},}\ }\href {\doibase
  10.1038/s41534-022-00594-4} {\bibfield  {journal} {\bibinfo  {journal} {npj
  Quantum Inf.}\ }\textbf {\bibinfo {volume} {8}},\ \bibinfo {pages} {88}
  (\bibinfo {year} {2022})}\BibitemShut {NoStop}%
\bibitem [{\citenamefont {Banaszek}\ and\ \citenamefont
  {W\'odkiewicz}(1998)}]{WignerEPR}%
  \BibitemOpen
  \bibfield  {author} {\bibinfo {author} {\bibfnamefont {K.}~\bibnamefont
  {Banaszek}}\ and\ \bibinfo {author} {\bibfnamefont {K.}~\bibnamefont
  {W\'odkiewicz}},\ }\bibfield  {title} {\enquote {\bibinfo {title}
  {Nonlocality of the einstein-podolsky-rosen state in the wigner
  representation},}\ }\href {\doibase 10.1103/PhysRevA.58.4345} {\bibfield
  {journal} {\bibinfo  {journal} {Phys. Rev. A}\ }\textbf {\bibinfo {volume}
  {58}},\ \bibinfo {pages} {4345--4347} (\bibinfo {year} {1998})}\BibitemShut
  {NoStop}%
\bibitem [{\citenamefont {Banaszek}\ \emph {et~al.}(1999)\citenamefont
  {Banaszek}, \citenamefont {Radzewicz}, \citenamefont {W\'odkiewicz},\ and\
  \citenamefont {Krasi\ifmmode~\acute{n}\else \'{n}\fi{}ski}}]{DirectWigner}%
  \BibitemOpen
  \bibfield  {author} {\bibinfo {author} {\bibfnamefont {K.}~\bibnamefont
  {Banaszek}}, \bibinfo {author} {\bibfnamefont {C.}~\bibnamefont {Radzewicz}},
  \bibinfo {author} {\bibfnamefont {K.}~\bibnamefont {W\'odkiewicz}}, \ and\
  \bibinfo {author} {\bibfnamefont {J.~S.}\ \bibnamefont
  {Krasi\ifmmode~\acute{n}\else \'{n}\fi{}ski}},\ }\bibfield  {title} {\enquote
  {\bibinfo {title} {Direct measurement of the wigner function by photon
  counting},}\ }\href {\doibase 10.1103/PhysRevA.60.674} {\bibfield  {journal}
  {\bibinfo  {journal} {Phys. Rev. A}\ }\textbf {\bibinfo {volume} {60}},\
  \bibinfo {pages} {674--677} (\bibinfo {year} {1999})}\BibitemShut {NoStop}%
\bibitem [{\citenamefont {Banaszek}\ and\ \citenamefont
  {W\'odkiewicz}(1996)}]{WignerPorbingPhase}%
  \BibitemOpen
  \bibfield  {author} {\bibinfo {author} {\bibfnamefont {K.}~\bibnamefont
  {Banaszek}}\ and\ \bibinfo {author} {\bibfnamefont {K.}~\bibnamefont
  {W\'odkiewicz}},\ }\bibfield  {title} {\enquote {\bibinfo {title} {Direct
  probing of quantum phase space by photon counting},}\ }\href {\doibase
  10.1103/PhysRevLett.76.4344} {\bibfield  {journal} {\bibinfo  {journal}
  {Phys. Rev. Lett.}\ }\textbf {\bibinfo {volume} {76}},\ \bibinfo {pages}
  {4344--4347} (\bibinfo {year} {1996})}\BibitemShut {NoStop}%
\bibitem [{\citenamefont {Besse}\ \emph {et~al.}(2020)\citenamefont {Besse},
  \citenamefont {Gasparinetti}, \citenamefont {Collodo}, \citenamefont
  {Walter}, \citenamefont {Remm}, \citenamefont {Krause}, \citenamefont
  {Eichler},\ and\ \citenamefont {Wallraff}}]{ParityDetectionOfField}%
  \BibitemOpen
  \bibfield  {author} {\bibinfo {author} {\bibfnamefont {J.-C.}\ \bibnamefont
  {Besse}}, \bibinfo {author} {\bibfnamefont {S.}~\bibnamefont {Gasparinetti}},
  \bibinfo {author} {\bibfnamefont {M.~C.}\ \bibnamefont {Collodo}}, \bibinfo
  {author} {\bibfnamefont {T.}~\bibnamefont {Walter}}, \bibinfo {author}
  {\bibfnamefont {A.}~\bibnamefont {Remm}}, \bibinfo {author} {\bibfnamefont
  {J.}~\bibnamefont {Krause}}, \bibinfo {author} {\bibfnamefont
  {C.}~\bibnamefont {Eichler}}, \ and\ \bibinfo {author} {\bibfnamefont
  {A.}~\bibnamefont {Wallraff}},\ }\bibfield  {title} {\enquote {\bibinfo
  {title} {Parity detection of propagating microwave fields},}\ }\href
  {\doibase 10.1103/PhysRevX.10.011046} {\bibfield  {journal} {\bibinfo
  {journal} {Phys. Rev. X}\ }\textbf {\bibinfo {volume} {10}},\ \bibinfo
  {pages} {011046} (\bibinfo {year} {2020})}\BibitemShut {NoStop}%
\bibitem [{\citenamefont {Lachance-Quirion}\ \emph {et~al.}(2019)\citenamefont
  {Lachance-Quirion}, \citenamefont {Tabuchi}, \citenamefont {Gloppe},
  \citenamefont {Usami},\ and\ \citenamefont {Nakamura}}]{HybridSystem}%
  \BibitemOpen
  \bibfield  {author} {\bibinfo {author} {\bibfnamefont {D.}~\bibnamefont
  {Lachance-Quirion}}, \bibinfo {author} {\bibfnamefont {Y.}~\bibnamefont
  {Tabuchi}}, \bibinfo {author} {\bibfnamefont {A.}~\bibnamefont {Gloppe}},
  \bibinfo {author} {\bibfnamefont {K.}~\bibnamefont {Usami}}, \ and\ \bibinfo
  {author} {\bibfnamefont {Y.}~\bibnamefont {Nakamura}},\ }\bibfield  {title}
  {\enquote {\bibinfo {title} {Hybrid quantum systems based on magnonics},}\
  }\href {\doibase 10.7567/1882-0786/ab248d} {\bibfield  {journal} {\bibinfo
  {journal} {Appl. Phys. Express}\ }\textbf {\bibinfo {volume} {12}},\ \bibinfo
  {pages} {070101} (\bibinfo {year} {2019})}\BibitemShut {NoStop}%
\bibitem [{\citenamefont {Yuan}\ \emph {et~al.}(2022)\citenamefont {Yuan},
  \citenamefont {Cao}, \citenamefont {Kamra}, \citenamefont {Duine},\ and\
  \citenamefont {Yan}}]{QuantumMagnonics}%
  \BibitemOpen
  \bibfield  {author} {\bibinfo {author} {\bibfnamefont {H.}~\bibnamefont
  {Yuan}}, \bibinfo {author} {\bibfnamefont {Y.}~\bibnamefont {Cao}}, \bibinfo
  {author} {\bibfnamefont {A.}~\bibnamefont {Kamra}}, \bibinfo {author}
  {\bibfnamefont {R.~A.}\ \bibnamefont {Duine}}, \ and\ \bibinfo {author}
  {\bibfnamefont {P.}~\bibnamefont {Yan}},\ }\bibfield  {title} {\enquote
  {\bibinfo {title} {Quantum magnonics: When magnon spintronics meets quantum
  information science},}\ }\href {\doibase
  https://doi.org/10.1016/j.physrep.2022.03.002} {\bibfield  {journal}
  {\bibinfo  {journal} {Phys. Rep.}\ }\textbf {\bibinfo {volume} {965}},\
  \bibinfo {pages} {1--74} (\bibinfo {year} {2022})}\BibitemShut {NoStop}%
\bibitem [{\citenamefont {{Zare Rameshti}}\ \emph {et~al.}(2022)\citenamefont
  {{Zare Rameshti}}, \citenamefont {{Viola Kusminskiy}}, \citenamefont {Haigh},
  \citenamefont {Usami}, \citenamefont {Lachance-Quirion}, \citenamefont
  {Nakamura}, \citenamefont {Hu}, \citenamefont {Tang}, \citenamefont {Bauer},\
  and\ \citenamefont {Blanter}}]{CavityMagnonics}%
  \BibitemOpen
  \bibfield  {author} {\bibinfo {author} {\bibfnamefont {B.}~\bibnamefont
  {{Zare Rameshti}}}, \bibinfo {author} {\bibfnamefont {S.}~\bibnamefont
  {{Viola Kusminskiy}}}, \bibinfo {author} {\bibfnamefont {J.~A.}\ \bibnamefont
  {Haigh}}, \bibinfo {author} {\bibfnamefont {K.}~\bibnamefont {Usami}},
  \bibinfo {author} {\bibfnamefont {D.}~\bibnamefont {Lachance-Quirion}},
  \bibinfo {author} {\bibfnamefont {Y.}~\bibnamefont {Nakamura}}, \bibinfo
  {author} {\bibfnamefont {C.-M.}\ \bibnamefont {Hu}}, \bibinfo {author}
  {\bibfnamefont {H.~X.}\ \bibnamefont {Tang}}, \bibinfo {author}
  {\bibfnamefont {G.~E.}\ \bibnamefont {Bauer}}, \ and\ \bibinfo {author}
  {\bibfnamefont {Y.~M.}\ \bibnamefont {Blanter}},\ }\bibfield  {title}
  {\enquote {\bibinfo {title} {Cavity magnonics},}\ }\href {\doibase
  https://doi.org/10.1016/j.physrep.2022.06.001} {\bibfield  {journal}
  {\bibinfo  {journal} {Phys. Rep.}\ }\textbf {\bibinfo {volume} {979}},\
  \bibinfo {pages} {1--61} (\bibinfo {year} {2022})}\BibitemShut {NoStop}%
\bibitem [{\citenamefont {Huebl}\ \emph {et~al.}(2013)\citenamefont {Huebl},
  \citenamefont {Zollitsch}, \citenamefont {Lotze}, \citenamefont {Hocke},
  \citenamefont {Greifenstein}, \citenamefont {Marx}, \citenamefont {Gross},\
  and\ \citenamefont {Goennenwein}}]{Cooperativity}%
  \BibitemOpen
  \bibfield  {author} {\bibinfo {author} {\bibfnamefont {H.}~\bibnamefont
  {Huebl}}, \bibinfo {author} {\bibfnamefont {C.~W.}\ \bibnamefont
  {Zollitsch}}, \bibinfo {author} {\bibfnamefont {J.}~\bibnamefont {Lotze}},
  \bibinfo {author} {\bibfnamefont {F.}~\bibnamefont {Hocke}}, \bibinfo
  {author} {\bibfnamefont {M.}~\bibnamefont {Greifenstein}}, \bibinfo {author}
  {\bibfnamefont {A.}~\bibnamefont {Marx}}, \bibinfo {author} {\bibfnamefont
  {R.}~\bibnamefont {Gross}}, \ and\ \bibinfo {author} {\bibfnamefont
  {S.~T.~B.}\ \bibnamefont {Goennenwein}},\ }\bibfield  {title} {\enquote
  {\bibinfo {title} {High cooperativity in coupled microwave resonator
  ferrimagnetic insulator hybrids},}\ }\href {\doibase
  10.1103/PhysRevLett.111.127003} {\bibfield  {journal} {\bibinfo  {journal}
  {Phys. Rev. Lett.}\ }\textbf {\bibinfo {volume} {111}},\ \bibinfo {pages}
  {127003} (\bibinfo {year} {2013})}\BibitemShut {NoStop}%
\bibitem [{\citenamefont {Tabuchi}\ \emph {et~al.}(2014)\citenamefont
  {Tabuchi}, \citenamefont {Ishino}, \citenamefont {Ishikawa}, \citenamefont
  {Yamazaki}, \citenamefont {Usami},\ and\ \citenamefont
  {Nakamura}}]{HybridizingMagnonsAndPhotons}%
  \BibitemOpen
  \bibfield  {author} {\bibinfo {author} {\bibfnamefont {Y.}~\bibnamefont
  {Tabuchi}}, \bibinfo {author} {\bibfnamefont {S.}~\bibnamefont {Ishino}},
  \bibinfo {author} {\bibfnamefont {T.}~\bibnamefont {Ishikawa}}, \bibinfo
  {author} {\bibfnamefont {R.}~\bibnamefont {Yamazaki}}, \bibinfo {author}
  {\bibfnamefont {K.}~\bibnamefont {Usami}}, \ and\ \bibinfo {author}
  {\bibfnamefont {Y.}~\bibnamefont {Nakamura}},\ }\bibfield  {title} {\enquote
  {\bibinfo {title} {Hybridizing ferromagnetic magnons and microwave photons in
  the quantum limit},}\ }\href {\doibase 10.1103/PhysRevLett.113.083603}
  {\bibfield  {journal} {\bibinfo  {journal} {Phys. Rev. Lett.}\ }\textbf
  {\bibinfo {volume} {113}},\ \bibinfo {pages} {083603} (\bibinfo {year}
  {2014})}\BibitemShut {NoStop}%
\bibitem [{\citenamefont {Zhang}\ \emph {et~al.}(2014)\citenamefont {Zhang},
  \citenamefont {Zou}, \citenamefont {Jiang},\ and\ \citenamefont
  {Tang}}]{PhotonsMagnonStrongCoupling}%
  \BibitemOpen
  \bibfield  {author} {\bibinfo {author} {\bibfnamefont {X.}~\bibnamefont
  {Zhang}}, \bibinfo {author} {\bibfnamefont {C.-L.}\ \bibnamefont {Zou}},
  \bibinfo {author} {\bibfnamefont {L.}~\bibnamefont {Jiang}}, \ and\ \bibinfo
  {author} {\bibfnamefont {H.~X.}\ \bibnamefont {Tang}},\ }\bibfield  {title}
  {\enquote {\bibinfo {title} {Strongly coupled magnons and cavity microwave
  photons},}\ }\href {\doibase 10.1103/PhysRevLett.113.156401} {\bibfield
  {journal} {\bibinfo  {journal} {Phys. Rev. Lett.}\ }\textbf {\bibinfo
  {volume} {113}},\ \bibinfo {pages} {156401} (\bibinfo {year}
  {2014})}\BibitemShut {NoStop}%
\bibitem [{\citenamefont {Bai}\ \emph {et~al.}(2015)\citenamefont {Bai},
  \citenamefont {Harder}, \citenamefont {Chen}, \citenamefont {Fan},
  \citenamefont {Xiao},\ and\ \citenamefont {Hu}}]{SpinPumping}%
  \BibitemOpen
  \bibfield  {author} {\bibinfo {author} {\bibfnamefont {L.}~\bibnamefont
  {Bai}}, \bibinfo {author} {\bibfnamefont {M.}~\bibnamefont {Harder}},
  \bibinfo {author} {\bibfnamefont {Y.~P.}\ \bibnamefont {Chen}}, \bibinfo
  {author} {\bibfnamefont {X.}~\bibnamefont {Fan}}, \bibinfo {author}
  {\bibfnamefont {J.~Q.}\ \bibnamefont {Xiao}}, \ and\ \bibinfo {author}
  {\bibfnamefont {C.-M.}\ \bibnamefont {Hu}},\ }\bibfield  {title} {\enquote
  {\bibinfo {title} {Spin pumping in electrodynamically coupled magnon-photon
  systems},}\ }\href {\doibase 10.1103/PhysRevLett.114.227201} {\bibfield
  {journal} {\bibinfo  {journal} {Phys. Rev. Lett.}\ }\textbf {\bibinfo
  {volume} {114}},\ \bibinfo {pages} {227201} (\bibinfo {year}
  {2015})}\BibitemShut {NoStop}%
\bibitem [{\citenamefont {Yu}, \citenamefont {Shen},\ and\ \citenamefont
  {Li}(2020)}]{PhotonPhotonByMagnon}%
  \BibitemOpen
  \bibfield  {author} {\bibinfo {author} {\bibfnamefont {M.}~\bibnamefont
  {Yu}}, \bibinfo {author} {\bibfnamefont {H.}~\bibnamefont {Shen}}, \ and\
  \bibinfo {author} {\bibfnamefont {J.}~\bibnamefont {Li}},\ }\bibfield
  {title} {\enquote {\bibinfo {title} {Magnetostrictively induced stationary
  entanglement between two microwave fields},}\ }\href {\doibase
  10.1103/PhysRevLett.124.213604} {\bibfield  {journal} {\bibinfo  {journal}
  {Phys. Rev. Lett.}\ }\textbf {\bibinfo {volume} {124}},\ \bibinfo {pages}
  {213604} (\bibinfo {year} {2020})}\BibitemShut {NoStop}%
\bibitem [{\citenamefont {Zhang}\ \emph {et~al.}(2016)\citenamefont {Zhang},
  \citenamefont {Zou}, \citenamefont {Jiang},\ and\ \citenamefont
  {Tang}}]{CavityMagnomechanics}%
  \BibitemOpen
  \bibfield  {author} {\bibinfo {author} {\bibfnamefont {X.}~\bibnamefont
  {Zhang}}, \bibinfo {author} {\bibfnamefont {C.-L.}\ \bibnamefont {Zou}},
  \bibinfo {author} {\bibfnamefont {L.}~\bibnamefont {Jiang}}, \ and\ \bibinfo
  {author} {\bibfnamefont {H.~X.}\ \bibnamefont {Tang}},\ }\bibfield  {title}
  {\enquote {\bibinfo {title} {Cavity magnomechanics},}\ }\href {\doibase
  10.1126/sciadv.1501286} {\bibfield  {journal} {\bibinfo  {journal} {Sci.
  Adv.}\ }\textbf {\bibinfo {volume} {2}},\ \bibinfo {pages} {e1501286}
  (\bibinfo {year} {2016})}\BibitemShut {NoStop}%
\bibitem [{\citenamefont {Potts}\ \emph {et~al.}(2021)\citenamefont {Potts},
  \citenamefont {Varga}, \citenamefont {Bittencourt}, \citenamefont
  {Kusminskiy},\ and\ \citenamefont
  {Davis}}]{DynamicalBackactionMagnomechanics}%
  \BibitemOpen
  \bibfield  {author} {\bibinfo {author} {\bibfnamefont {C.~A.}\ \bibnamefont
  {Potts}}, \bibinfo {author} {\bibfnamefont {E.}~\bibnamefont {Varga}},
  \bibinfo {author} {\bibfnamefont {V.~A. S.~V.}\ \bibnamefont {Bittencourt}},
  \bibinfo {author} {\bibfnamefont {S.~V.}\ \bibnamefont {Kusminskiy}}, \ and\
  \bibinfo {author} {\bibfnamefont {J.~P.}\ \bibnamefont {Davis}},\ }\bibfield
  {title} {\enquote {\bibinfo {title} {Dynamical backaction magnomechanics},}\
  }\href {\doibase 10.1103/PhysRevX.11.031053} {\bibfield  {journal} {\bibinfo
  {journal} {Phys. Rev. X}\ }\textbf {\bibinfo {volume} {11}},\ \bibinfo
  {pages} {031053} (\bibinfo {year} {2021})}\BibitemShut {NoStop}%
\bibitem [{\citenamefont {Tabuchi}\ \emph {et~al.}(2015)\citenamefont
  {Tabuchi}, \citenamefont {Ishino}, \citenamefont {Noguchi}, \citenamefont
  {Ishikawa}, \citenamefont {Yamazaki}, \citenamefont {Usami},\ and\
  \citenamefont {Nakamura}}]{MagnonQubitCoupling}%
  \BibitemOpen
  \bibfield  {author} {\bibinfo {author} {\bibfnamefont {Y.}~\bibnamefont
  {Tabuchi}}, \bibinfo {author} {\bibfnamefont {S.}~\bibnamefont {Ishino}},
  \bibinfo {author} {\bibfnamefont {A.}~\bibnamefont {Noguchi}}, \bibinfo
  {author} {\bibfnamefont {T.}~\bibnamefont {Ishikawa}}, \bibinfo {author}
  {\bibfnamefont {R.}~\bibnamefont {Yamazaki}}, \bibinfo {author}
  {\bibfnamefont {K.}~\bibnamefont {Usami}}, \ and\ \bibinfo {author}
  {\bibfnamefont {Y.}~\bibnamefont {Nakamura}},\ }\bibfield  {title} {\enquote
  {\bibinfo {title} {Coherent coupling between a ferromagnetic magnon and a
  superconducting qubit},}\ }\href {\doibase 10.1126/science.aaa3693}
  {\bibfield  {journal} {\bibinfo  {journal} {Science}\ }\textbf {\bibinfo
  {volume} {349}},\ \bibinfo {pages} {405--408} (\bibinfo {year}
  {2015})}\BibitemShut {NoStop}%
\bibitem [{\citenamefont {Tabuchi}\ \emph {et~al.}(2016)\citenamefont
  {Tabuchi}, \citenamefont {Ishino}, \citenamefont {Noguchi}, \citenamefont
  {Ishikawa}, \citenamefont {Yamazaki}, \citenamefont {Usami},\ and\
  \citenamefont {Nakamura}}]{MagnonMeetQubit}%
  \BibitemOpen
  \bibfield  {author} {\bibinfo {author} {\bibfnamefont {Y.}~\bibnamefont
  {Tabuchi}}, \bibinfo {author} {\bibfnamefont {S.}~\bibnamefont {Ishino}},
  \bibinfo {author} {\bibfnamefont {A.}~\bibnamefont {Noguchi}}, \bibinfo
  {author} {\bibfnamefont {T.}~\bibnamefont {Ishikawa}}, \bibinfo {author}
  {\bibfnamefont {R.}~\bibnamefont {Yamazaki}}, \bibinfo {author}
  {\bibfnamefont {K.}~\bibnamefont {Usami}}, \ and\ \bibinfo {author}
  {\bibfnamefont {Y.}~\bibnamefont {Nakamura}},\ }\bibfield  {title} {\enquote
  {\bibinfo {title} {Quantum magnonics: The magnon meets the superconducting
  qubit},}\ }\href {\doibase https://doi.org/10.1016/j.crhy.2016.07.009}
  {\bibfield  {journal} {\bibinfo  {journal} {C. R. Phys.}\ }\textbf {\bibinfo
  {volume} {17}},\ \bibinfo {pages} {729--739} (\bibinfo {year}
  {2016})}\BibitemShut {NoStop}%
\bibitem [{\citenamefont {Lachance-Quirion}\ \emph {et~al.}(2017)\citenamefont
  {Lachance-Quirion}, \citenamefont {Tabuchi}, \citenamefont {Ishino},
  \citenamefont {Noguchi}, \citenamefont {Ishikawa}, \citenamefont {Yamazaki},\
  and\ \citenamefont {Nakamura}}]{ResolvingQuanta}%
  \BibitemOpen
  \bibfield  {author} {\bibinfo {author} {\bibfnamefont {D.}~\bibnamefont
  {Lachance-Quirion}}, \bibinfo {author} {\bibfnamefont {Y.}~\bibnamefont
  {Tabuchi}}, \bibinfo {author} {\bibfnamefont {S.}~\bibnamefont {Ishino}},
  \bibinfo {author} {\bibfnamefont {A.}~\bibnamefont {Noguchi}}, \bibinfo
  {author} {\bibfnamefont {T.}~\bibnamefont {Ishikawa}}, \bibinfo {author}
  {\bibfnamefont {R.}~\bibnamefont {Yamazaki}}, \ and\ \bibinfo {author}
  {\bibfnamefont {Y.}~\bibnamefont {Nakamura}},\ }\bibfield  {title} {\enquote
  {\bibinfo {title} {Resolving quanta of collective spin excitations in a
  millimeter-sized ferromagnet},}\ }\href {\doibase 10.1126/sciadv.1603150}
  {\bibfield  {journal} {\bibinfo  {journal} {Sci. Adv.}\ }\textbf {\bibinfo
  {volume} {3}},\ \bibinfo {pages} {e1603150} (\bibinfo {year}
  {2017})}\BibitemShut {NoStop}%
\bibitem [{\citenamefont {Nair}\ and\ \citenamefont
  {Agarwal}(2020)}]{EntanglementFerriteSamples}%
  \BibitemOpen
  \bibfield  {author} {\bibinfo {author} {\bibfnamefont {J.~M.~P.}\
  \bibnamefont {Nair}}\ and\ \bibinfo {author} {\bibfnamefont {G.~S.}\
  \bibnamefont {Agarwal}},\ }\bibfield  {title} {\enquote {\bibinfo {title}
  {{Deterministic quantum entanglement between macroscopic ferrite samples}},}\
  }\href {\doibase 10.1063/5.0015195} {\bibfield  {journal} {\bibinfo
  {journal} {Appl. Phys. Lett.}\ }\textbf {\bibinfo {volume} {117}},\ \bibinfo
  {pages} {084001} (\bibinfo {year} {2020})}\BibitemShut {NoStop}%
\bibitem [{\citenamefont {Yu}, \citenamefont {Zhu},\ and\ \citenamefont
  {Li}(2020)}]{TwoMagnonEntanglement}%
  \BibitemOpen
  \bibfield  {author} {\bibinfo {author} {\bibfnamefont {M.}~\bibnamefont
  {Yu}}, \bibinfo {author} {\bibfnamefont {S.-Y.}\ \bibnamefont {Zhu}}, \ and\
  \bibinfo {author} {\bibfnamefont {J.}~\bibnamefont {Li}},\ }\bibfield
  {title} {\enquote {\bibinfo {title} {Macroscopic entanglement of two magnon
  modes via quantum correlated microwave fields},}\ }\href {\doibase
  10.1088/1361-6455/ab68b5} {\bibfield  {journal} {\bibinfo  {journal} {J.
  Phys. B: At., Mol., Opt. Phys.}\ }\textbf {\bibinfo {volume} {53}},\ \bibinfo
  {pages} {065402} (\bibinfo {year} {2020})}\BibitemShut {NoStop}%
\bibitem [{\citenamefont {Zhang}, \citenamefont {Scully},\ and\ \citenamefont
  {Agarwal}(2019)}]{KerrMagnonEntanglement}%
  \BibitemOpen
  \bibfield  {author} {\bibinfo {author} {\bibfnamefont {Z.}~\bibnamefont
  {Zhang}}, \bibinfo {author} {\bibfnamefont {M.~O.}\ \bibnamefont {Scully}}, \
  and\ \bibinfo {author} {\bibfnamefont {G.~S.}\ \bibnamefont {Agarwal}},\
  }\bibfield  {title} {\enquote {\bibinfo {title} {Quantum entanglement between
  two magnon modes via kerr nonlinearity driven far from equilibrium},}\ }\href
  {\doibase 10.1103/PhysRevResearch.1.023021} {\bibfield  {journal} {\bibinfo
  {journal} {Phys. Rev. Res.}\ }\textbf {\bibinfo {volume} {1}},\ \bibinfo
  {pages} {023021} (\bibinfo {year} {2019})}\BibitemShut {NoStop}%
\bibitem [{\citenamefont {Li}\ and\ \citenamefont
  {Zhu}(2019)}]{MagnetostrictiveMagnonEntanglement}%
  \BibitemOpen
  \bibfield  {author} {\bibinfo {author} {\bibfnamefont {J.}~\bibnamefont
  {Li}}\ and\ \bibinfo {author} {\bibfnamefont {S.-Y.}\ \bibnamefont {Zhu}},\
  }\bibfield  {title} {\enquote {\bibinfo {title} {Entangling two magnon modes
  via magnetostrictive interaction},}\ }\href {\doibase
  10.1088/1367-2630/ab3508} {\bibfield  {journal} {\bibinfo  {journal} {New J.
  Phys.}\ }\textbf {\bibinfo {volume} {21}},\ \bibinfo {pages} {085001}
  (\bibinfo {year} {2019})}\BibitemShut {NoStop}%
\bibitem [{\citenamefont {Wu}\ \emph {et~al.}(2021)\citenamefont {Wu},
  \citenamefont {Wang}, \citenamefont {Wu}, \citenamefont {Li},\ and\
  \citenamefont {You}}]{RemoteMagnonEntanglement}%
  \BibitemOpen
  \bibfield  {author} {\bibinfo {author} {\bibfnamefont {W.-J.}\ \bibnamefont
  {Wu}}, \bibinfo {author} {\bibfnamefont {Y.-P.}\ \bibnamefont {Wang}},
  \bibinfo {author} {\bibfnamefont {J.-Z.}\ \bibnamefont {Wu}}, \bibinfo
  {author} {\bibfnamefont {J.}~\bibnamefont {Li}}, \ and\ \bibinfo {author}
  {\bibfnamefont {J.~Q.}\ \bibnamefont {You}},\ }\bibfield  {title} {\enquote
  {\bibinfo {title} {Remote magnon entanglement between two massive
  ferrimagnetic spheres via cavity optomagnonics},}\ }\href {\doibase
  10.1103/PhysRevA.104.023711} {\bibfield  {journal} {\bibinfo  {journal}
  {Phys. Rev. A}\ }\textbf {\bibinfo {volume} {104}},\ \bibinfo {pages}
  {023711} (\bibinfo {year} {2021})}\BibitemShut {NoStop}%
\bibitem [{\citenamefont {Yuan}\ \emph {et~al.}(2020)\citenamefont {Yuan},
  \citenamefont {Zheng}, \citenamefont {Ficek}, \citenamefont {He},\ and\
  \citenamefont {Yung}}]{EnhanceMagnonMagnon}%
  \BibitemOpen
  \bibfield  {author} {\bibinfo {author} {\bibfnamefont {H.~Y.}\ \bibnamefont
  {Yuan}}, \bibinfo {author} {\bibfnamefont {S.}~\bibnamefont {Zheng}},
  \bibinfo {author} {\bibfnamefont {Z.}~\bibnamefont {Ficek}}, \bibinfo
  {author} {\bibfnamefont {Q.~Y.}\ \bibnamefont {He}}, \ and\ \bibinfo {author}
  {\bibfnamefont {M.-H.}\ \bibnamefont {Yung}},\ }\bibfield  {title} {\enquote
  {\bibinfo {title} {Enhancement of magnon-magnon entanglement inside a
  cavity},}\ }\href {\doibase 10.1103/PhysRevB.101.014419} {\bibfield
  {journal} {\bibinfo  {journal} {Phys. Rev. B}\ }\textbf {\bibinfo {volume}
  {101}},\ \bibinfo {pages} {014419} (\bibinfo {year} {2020})}\BibitemShut
  {NoStop}%
\bibitem [{\citenamefont {Azimi~Mousolou}\ \emph {et~al.}(2021)\citenamefont
  {Azimi~Mousolou}, \citenamefont {Liu}, \citenamefont {Bergman}, \citenamefont
  {Delin}, \citenamefont {Eriksson}, \citenamefont {Pereiro}, \citenamefont
  {Thonig},\ and\ \citenamefont {Sj\"oqvist}}]{MagnonMagnonInCavity}%
  \BibitemOpen
  \bibfield  {author} {\bibinfo {author} {\bibfnamefont {V.}~\bibnamefont
  {Azimi~Mousolou}}, \bibinfo {author} {\bibfnamefont {Y.}~\bibnamefont {Liu}},
  \bibinfo {author} {\bibfnamefont {A.}~\bibnamefont {Bergman}}, \bibinfo
  {author} {\bibfnamefont {A.}~\bibnamefont {Delin}}, \bibinfo {author}
  {\bibfnamefont {O.}~\bibnamefont {Eriksson}}, \bibinfo {author}
  {\bibfnamefont {M.}~\bibnamefont {Pereiro}}, \bibinfo {author} {\bibfnamefont
  {D.}~\bibnamefont {Thonig}}, \ and\ \bibinfo {author} {\bibfnamefont
  {E.}~\bibnamefont {Sj\"oqvist}},\ }\bibfield  {title} {\enquote {\bibinfo
  {title} {Magnon-magnon entanglement and its quantification via a microwave
  cavity},}\ }\href {\doibase 10.1103/PhysRevB.104.224302} {\bibfield
  {journal} {\bibinfo  {journal} {Phys. Rev. B}\ }\textbf {\bibinfo {volume}
  {104}},\ \bibinfo {pages} {224302} (\bibinfo {year} {2021})}\BibitemShut
  {NoStop}%
\bibitem [{\citenamefont {Daiss}\ \emph {et~al.}(2019)\citenamefont {Daiss},
  \citenamefont {Welte}, \citenamefont {Hacker}, \citenamefont {Li},\ and\
  \citenamefont {Rempe}}]{SinglePhotonByParityProjection}%
  \BibitemOpen
  \bibfield  {author} {\bibinfo {author} {\bibfnamefont {S.}~\bibnamefont
  {Daiss}}, \bibinfo {author} {\bibfnamefont {S.}~\bibnamefont {Welte}},
  \bibinfo {author} {\bibfnamefont {B.}~\bibnamefont {Hacker}}, \bibinfo
  {author} {\bibfnamefont {L.}~\bibnamefont {Li}}, \ and\ \bibinfo {author}
  {\bibfnamefont {G.}~\bibnamefont {Rempe}},\ }\bibfield  {title} {\enquote
  {\bibinfo {title} {Single-photon distillation via a photonic parity
  measurement using cavity qed},}\ }\href {\doibase
  10.1103/PhysRevLett.122.133603} {\bibfield  {journal} {\bibinfo  {journal}
  {Phys. Rev. Lett.}\ }\textbf {\bibinfo {volume} {122}},\ \bibinfo {pages}
  {133603} (\bibinfo {year} {2019})}\BibitemShut {NoStop}%
\bibitem [{\citenamefont {Xu}\ \emph {et~al.}(2023{\natexlab{a}})\citenamefont
  {Xu}, \citenamefont {Gu}, \citenamefont {Li}, \citenamefont {Weng},
  \citenamefont {Wang}, \citenamefont {Li}, \citenamefont {Wang}, \citenamefont
  {Zhu},\ and\ \citenamefont {You}}]{MagnonXu}%
  \BibitemOpen
  \bibfield  {author} {\bibinfo {author} {\bibfnamefont {D.}~\bibnamefont
  {Xu}}, \bibinfo {author} {\bibfnamefont {X.-K.}\ \bibnamefont {Gu}}, \bibinfo
  {author} {\bibfnamefont {H.-K.}\ \bibnamefont {Li}}, \bibinfo {author}
  {\bibfnamefont {Y.-C.}\ \bibnamefont {Weng}}, \bibinfo {author}
  {\bibfnamefont {Y.-P.}\ \bibnamefont {Wang}}, \bibinfo {author}
  {\bibfnamefont {J.}~\bibnamefont {Li}}, \bibinfo {author} {\bibfnamefont
  {H.}~\bibnamefont {Wang}}, \bibinfo {author} {\bibfnamefont {S.-Y.}\
  \bibnamefont {Zhu}}, \ and\ \bibinfo {author} {\bibfnamefont {J.~Q.}\
  \bibnamefont {You}},\ }\bibfield  {title} {\enquote {\bibinfo {title}
  {Quantum control of a single magnon in a macroscopic spin system},}\ }\href
  {\doibase 10.1103/PhysRevLett.130.193603} {\bibfield  {journal} {\bibinfo
  {journal} {Phys. Rev. Lett.}\ }\textbf {\bibinfo {volume} {130}},\ \bibinfo
  {pages} {193603} (\bibinfo {year} {2023}{\natexlab{a}})}\BibitemShut
  {NoStop}%
\bibitem [{\citenamefont {Kong}\ \emph {et~al.}(2021)\citenamefont {Kong},
  \citenamefont {Hu}, \citenamefont {Hu},\ and\ \citenamefont
  {Xu}}]{EntanglementByDressedField}%
  \BibitemOpen
  \bibfield  {author} {\bibinfo {author} {\bibfnamefont {D.}~\bibnamefont
  {Kong}}, \bibinfo {author} {\bibfnamefont {X.}~\bibnamefont {Hu}}, \bibinfo
  {author} {\bibfnamefont {L.}~\bibnamefont {Hu}}, \ and\ \bibinfo {author}
  {\bibfnamefont {J.}~\bibnamefont {Xu}},\ }\bibfield  {title} {\enquote
  {\bibinfo {title} {Magnon-atom interaction via dispersive cavities: Magnon
  entanglement},}\ }\href {\doibase 10.1103/PhysRevB.103.224416} {\bibfield
  {journal} {\bibinfo  {journal} {Phys. Rev. B}\ }\textbf {\bibinfo {volume}
  {103}},\ \bibinfo {pages} {224416} (\bibinfo {year} {2021})}\BibitemShut
  {NoStop}%
\bibitem [{\citenamefont {Blais}\ \emph {et~al.}(2021)\citenamefont {Blais},
  \citenamefont {Grimsmo}, \citenamefont {Girvin},\ and\ \citenamefont
  {Wallraff}}]{CircuteQED}%
  \BibitemOpen
  \bibfield  {author} {\bibinfo {author} {\bibfnamefont {A.}~\bibnamefont
  {Blais}}, \bibinfo {author} {\bibfnamefont {A.~L.}\ \bibnamefont {Grimsmo}},
  \bibinfo {author} {\bibfnamefont {S.~M.}\ \bibnamefont {Girvin}}, \ and\
  \bibinfo {author} {\bibfnamefont {A.}~\bibnamefont {Wallraff}},\ }\bibfield
  {title} {\enquote {\bibinfo {title} {Circuit quantum electrodynamics},}\
  }\href {\doibase 10.1103/RevModPhys.93.025005} {\bibfield  {journal}
  {\bibinfo  {journal} {Rev. Mod. Phys.}\ }\textbf {\bibinfo {volume} {93}},\
  \bibinfo {pages} {025005} (\bibinfo {year} {2021})}\BibitemShut {NoStop}%
\bibitem [{\citenamefont {Schrieffer}\ and\ \citenamefont
  {Wolff}(1966)}]{SWtransformation}%
  \BibitemOpen
  \bibfield  {author} {\bibinfo {author} {\bibfnamefont {J.~R.}\ \bibnamefont
  {Schrieffer}}\ and\ \bibinfo {author} {\bibfnamefont {P.~A.}\ \bibnamefont
  {Wolff}},\ }\bibfield  {title} {\enquote {\bibinfo {title} {Relation between
  the anderson and kondo hamiltonians},}\ }\href {\doibase
  10.1103/PhysRev.149.491} {\bibfield  {journal} {\bibinfo  {journal} {Phys.
  Rev.}\ }\textbf {\bibinfo {volume} {149}},\ \bibinfo {pages} {491--492}
  (\bibinfo {year} {1966})}\BibitemShut {NoStop}%
\bibitem [{\citenamefont {Xu}\ \emph {et~al.}(2023{\natexlab{b}})\citenamefont
  {Xu}, \citenamefont {Gu}, \citenamefont {Weng}, \citenamefont {Li},
  \citenamefont {Wang}, \citenamefont {Zhu},\ and\ \citenamefont
  {You}}]{xuBellstate}%
  \BibitemOpen
  \bibfield  {author} {\bibinfo {author} {\bibfnamefont {D.}~\bibnamefont
  {Xu}}, \bibinfo {author} {\bibfnamefont {X.-K.}\ \bibnamefont {Gu}}, \bibinfo
  {author} {\bibfnamefont {Y.-C.}\ \bibnamefont {Weng}}, \bibinfo {author}
  {\bibfnamefont {H.-K.}\ \bibnamefont {Li}}, \bibinfo {author} {\bibfnamefont
  {Y.-P.}\ \bibnamefont {Wang}}, \bibinfo {author} {\bibfnamefont {S.-Y.}\
  \bibnamefont {Zhu}}, \ and\ \bibinfo {author} {\bibfnamefont {J.~Q.}\
  \bibnamefont {You}},\ }\bibfield  {title} {\enquote {\bibinfo {title}
  {Deterministic generation and tomography of a macroscopic bell state between
  a millimeter-sized spin system and a superconducting qubit},}\ }\href
  {https://arxiv.org/abs/2306.09677} {\bibfield  {journal} {\bibinfo  {journal}
  {arXiv}\ ,\ \bibinfo {pages} {2306.09677}} (\bibinfo {year}
  {2023}{\natexlab{b}})}\BibitemShut {NoStop}%
\bibitem [{\citenamefont {Thomas}\ \emph {et~al.}(2021)\citenamefont {Thomas},
  \citenamefont {Parniak}, \citenamefont {{\O}stfeldt}, \citenamefont
  {M{\o}ller}, \citenamefont {B{\ae}rentsen}, \citenamefont {Tsaturyan},
  \citenamefont {Schliesser}, \citenamefont {Appel}, \citenamefont {Zeuthen},\
  and\ \citenamefont {Polzik}}]{MechanicalAndSpinEntanglement}%
  \BibitemOpen
  \bibfield  {author} {\bibinfo {author} {\bibfnamefont {R.~A.}\ \bibnamefont
  {Thomas}}, \bibinfo {author} {\bibfnamefont {M.}~\bibnamefont {Parniak}},
  \bibinfo {author} {\bibfnamefont {C.}~\bibnamefont {{\O}stfeldt}}, \bibinfo
  {author} {\bibfnamefont {C.~B.}\ \bibnamefont {M{\o}ller}}, \bibinfo {author}
  {\bibfnamefont {C.}~\bibnamefont {B{\ae}rentsen}}, \bibinfo {author}
  {\bibfnamefont {Y.}~\bibnamefont {Tsaturyan}}, \bibinfo {author}
  {\bibfnamefont {A.}~\bibnamefont {Schliesser}}, \bibinfo {author}
  {\bibfnamefont {J.}~\bibnamefont {Appel}}, \bibinfo {author} {\bibfnamefont
  {E.}~\bibnamefont {Zeuthen}}, \ and\ \bibinfo {author} {\bibfnamefont
  {E.~S.}\ \bibnamefont {Polzik}},\ }\bibfield  {title} {\enquote {\bibinfo
  {title} {Entanglement between distant macroscopic mechanical and spin
  systems},}\ }\href {\doibase 10.1038/s41567-020-1031-5} {\bibfield  {journal}
  {\bibinfo  {journal} {Nat. Phys.}\ }\textbf {\bibinfo {volume} {17}},\
  \bibinfo {pages} {228--233} (\bibinfo {year} {2021})}\BibitemShut {NoStop}%
\bibitem [{\citenamefont {Kotler}\ \emph {et~al.}(2021)\citenamefont {Kotler},
  \citenamefont {Peterson}, \citenamefont {Shojaee}, \citenamefont {Lecocq},
  \citenamefont {Cicak}, \citenamefont {Kwiatkowski}, \citenamefont {Geller},
  \citenamefont {Glancy}, \citenamefont {Knill}, \citenamefont {Simmonds},
  \citenamefont {Aumentado},\ and\ \citenamefont
  {Teufel}}]{MacroscopicEntanglement}%
  \BibitemOpen
  \bibfield  {author} {\bibinfo {author} {\bibfnamefont {S.}~\bibnamefont
  {Kotler}}, \bibinfo {author} {\bibfnamefont {G.~A.}\ \bibnamefont
  {Peterson}}, \bibinfo {author} {\bibfnamefont {E.}~\bibnamefont {Shojaee}},
  \bibinfo {author} {\bibfnamefont {F.}~\bibnamefont {Lecocq}}, \bibinfo
  {author} {\bibfnamefont {K.}~\bibnamefont {Cicak}}, \bibinfo {author}
  {\bibfnamefont {A.}~\bibnamefont {Kwiatkowski}}, \bibinfo {author}
  {\bibfnamefont {S.}~\bibnamefont {Geller}}, \bibinfo {author} {\bibfnamefont
  {S.}~\bibnamefont {Glancy}}, \bibinfo {author} {\bibfnamefont
  {E.}~\bibnamefont {Knill}}, \bibinfo {author} {\bibfnamefont {R.~W.}\
  \bibnamefont {Simmonds}}, \bibinfo {author} {\bibfnamefont {J.}~\bibnamefont
  {Aumentado}}, \ and\ \bibinfo {author} {\bibfnamefont {J.~D.}\ \bibnamefont
  {Teufel}},\ }\bibfield  {title} {\enquote {\bibinfo {title} {Direct
  observation of deterministic macroscopic entanglement},}\ }\href {\doibase
  10.1126/science.abf2998} {\bibfield  {journal} {\bibinfo  {journal}
  {Science}\ }\textbf {\bibinfo {volume} {372}},\ \bibinfo {pages} {622--625}
  (\bibinfo {year} {2021})}\BibitemShut {NoStop}%
\bibitem [{\citenamefont {Puebla}, \citenamefont {Abah},\ and\ \citenamefont
  {Paternostro}(2020)}]{CoolingNonlinear}%
  \BibitemOpen
  \bibfield  {author} {\bibinfo {author} {\bibfnamefont {R.}~\bibnamefont
  {Puebla}}, \bibinfo {author} {\bibfnamefont {O.}~\bibnamefont {Abah}}, \ and\
  \bibinfo {author} {\bibfnamefont {M.}~\bibnamefont {Paternostro}},\
  }\bibfield  {title} {\enquote {\bibinfo {title} {Measurement-based cooling of
  a nonlinear mechanical resonator},}\ }\href {\doibase
  10.1103/PhysRevB.101.245410} {\bibfield  {journal} {\bibinfo  {journal}
  {Phys. Rev. B}\ }\textbf {\bibinfo {volume} {101}},\ \bibinfo {pages}
  {245410} (\bibinfo {year} {2020})}\BibitemShut {NoStop}%
\bibitem [{\citenamefont {Caneva}, \citenamefont {Calarco},\ and\ \citenamefont
  {Montangero}(2011)}]{CRABpra}%
  \BibitemOpen
  \bibfield  {author} {\bibinfo {author} {\bibfnamefont {T.}~\bibnamefont
  {Caneva}}, \bibinfo {author} {\bibfnamefont {T.}~\bibnamefont {Calarco}}, \
  and\ \bibinfo {author} {\bibfnamefont {S.}~\bibnamefont {Montangero}},\
  }\bibfield  {title} {\enquote {\bibinfo {title} {Chopped random-basis quantum
  optimization},}\ }\href {\doibase 10.1103/PhysRevA.84.022326} {\bibfield
  {journal} {\bibinfo  {journal} {Phys. Rev. A}\ }\textbf {\bibinfo {volume}
  {84}},\ \bibinfo {pages} {022326} (\bibinfo {year} {2011})}\BibitemShut
  {NoStop}%
\bibitem [{\citenamefont {Doria}, \citenamefont {Calarco},\ and\ \citenamefont
  {Montangero}(2011)}]{CRABprl}%
  \BibitemOpen
  \bibfield  {author} {\bibinfo {author} {\bibfnamefont {P.}~\bibnamefont
  {Doria}}, \bibinfo {author} {\bibfnamefont {T.}~\bibnamefont {Calarco}}, \
  and\ \bibinfo {author} {\bibfnamefont {S.}~\bibnamefont {Montangero}},\
  }\bibfield  {title} {\enquote {\bibinfo {title} {Optimal control technique
  for many-body quantum dynamics},}\ }\href {\doibase
  10.1103/PhysRevLett.106.190501} {\bibfield  {journal} {\bibinfo  {journal}
  {Phys. Rev. Lett.}\ }\textbf {\bibinfo {volume} {106}},\ \bibinfo {pages}
  {190501} (\bibinfo {year} {2011})}\BibitemShut {NoStop}%
\bibitem [{\citenamefont {Nelder}\ and\ \citenamefont
  {Mead}(1965)}]{NelderMead}%
  \BibitemOpen
  \bibfield  {author} {\bibinfo {author} {\bibfnamefont {J.~A.}\ \bibnamefont
  {Nelder}}\ and\ \bibinfo {author} {\bibfnamefont {R.}~\bibnamefont {Mead}},\
  }\bibfield  {title} {\enquote {\bibinfo {title} {{A Simplex Method for
  Function Minimization}},}\ }\href {\doibase 10.1093/comjnl/7.4.308}
  {\bibfield  {journal} {\bibinfo  {journal} {Comput. J.}\ }\textbf {\bibinfo
  {volume} {7}},\ \bibinfo {pages} {308--313} (\bibinfo {year}
  {1965})}\BibitemShut {NoStop}%
\bibitem [{\citenamefont {Sun}\ \emph {et~al.}(2021)\citenamefont {Sun},
  \citenamefont {Zheng}, \citenamefont {Xiao}, \citenamefont {Gong},
  \citenamefont {He},\ and\ \citenamefont {Xia}}]{CatStateHe}%
  \BibitemOpen
  \bibfield  {author} {\bibinfo {author} {\bibfnamefont {F.-X.}\ \bibnamefont
  {Sun}}, \bibinfo {author} {\bibfnamefont {S.-S.}\ \bibnamefont {Zheng}},
  \bibinfo {author} {\bibfnamefont {Y.}~\bibnamefont {Xiao}}, \bibinfo {author}
  {\bibfnamefont {Q.}~\bibnamefont {Gong}}, \bibinfo {author} {\bibfnamefont
  {Q.}~\bibnamefont {He}}, \ and\ \bibinfo {author} {\bibfnamefont
  {K.}~\bibnamefont {Xia}},\ }\bibfield  {title} {\enquote {\bibinfo {title}
  {Remote generation of magnon schr\"odinger cat state via magnon-photon
  entanglement},}\ }\href {\doibase 10.1103/PhysRevLett.127.087203} {\bibfield
  {journal} {\bibinfo  {journal} {Phys. Rev. Lett.}\ }\textbf {\bibinfo
  {volume} {127}},\ \bibinfo {pages} {087203} (\bibinfo {year}
  {2021})}\BibitemShut {NoStop}%
\bibitem [{\citenamefont {Pezz\`e}\ \emph {et~al.}(2019)\citenamefont
  {Pezz\`e}, \citenamefont {Gessner}, \citenamefont {Feldmann}, \citenamefont
  {Klempt}, \citenamefont {Santos},\ and\ \citenamefont
  {Smerzi}}]{BECSuperposition}%
  \BibitemOpen
  \bibfield  {author} {\bibinfo {author} {\bibfnamefont {L.}~\bibnamefont
  {Pezz\`e}}, \bibinfo {author} {\bibfnamefont {M.}~\bibnamefont {Gessner}},
  \bibinfo {author} {\bibfnamefont {P.}~\bibnamefont {Feldmann}}, \bibinfo
  {author} {\bibfnamefont {C.}~\bibnamefont {Klempt}}, \bibinfo {author}
  {\bibfnamefont {L.}~\bibnamefont {Santos}}, \ and\ \bibinfo {author}
  {\bibfnamefont {A.}~\bibnamefont {Smerzi}},\ }\bibfield  {title} {\enquote
  {\bibinfo {title} {Heralded generation of macroscopic superposition states in
  a spinor bose-einstein condensate},}\ }\href {\doibase
  10.1103/PhysRevLett.123.260403} {\bibfield  {journal} {\bibinfo  {journal}
  {Phys. Rev. Lett.}\ }\textbf {\bibinfo {volume} {123}},\ \bibinfo {pages}
  {260403} (\bibinfo {year} {2019})}\BibitemShut {NoStop}%
\bibitem [{\citenamefont {Guo}\ \emph {et~al.}(2023)\citenamefont {Guo},
  \citenamefont {Sun}, \citenamefont {He},\ and\ \citenamefont
  {Fadel}}]{SuperpositionHe}%
  \BibitemOpen
  \bibfield  {author} {\bibinfo {author} {\bibfnamefont {J.}~\bibnamefont
  {Guo}}, \bibinfo {author} {\bibfnamefont {F.}~\bibnamefont {Sun}}, \bibinfo
  {author} {\bibfnamefont {Q.}~\bibnamefont {He}}, \ and\ \bibinfo {author}
  {\bibfnamefont {M.}~\bibnamefont {Fadel}},\ }\bibfield  {title} {\enquote
  {\bibinfo {title} {Assisted metrology and preparation of macroscopic
  superpositions with split spin-squeezed states},}\ }\href {\doibase
  10.1103/PhysRevA.108.053327} {\bibfield  {journal} {\bibinfo  {journal}
  {Phys. Rev. A}\ }\textbf {\bibinfo {volume} {108}},\ \bibinfo {pages}
  {053327} (\bibinfo {year} {2023})}\BibitemShut {NoStop}%
\end{thebibliography}%

\end{document}